\documentclass[12pt]{article}
\usepackage{geometry} 
\usepackage{epstopdf}
\usepackage{epsfig}
\usepackage{amsmath}
\usepackage{booktabs}
\usepackage{longtable}
\usepackage{subfig}
\usepackage[hyphens]{url}
\usepackage{natbib,hyperref}        
\title{How Many Components should be Retained from a Multivariate Time Series PCA?}

\author{Alethea Rea$^{1}$ and William Rea$^{2}$, \\
1. Centre for Applied Statistics, University of Western Australia, Australia \\
2. Department of Economics and Finance, University of Canterbury, \\
New Zealand  }

\begin{document}

\maketitle

\begin{abstract}

We report on the results of two new approaches to considering how many
principal components to retain from an analysis of a multivariate 
time series. The first is by
using a ``heat map'' based approach.
A heat map in this context refers to a series of 
principal component coefficients created by applying a sliding window 
to a multivariate time series.  Furthermore the heat maps can provide
detailed insights into the evolution of the structure of each principal 
component over time. The second is by examining the change of the angle
of the principal component over time within the high-dimensional data space.
We provide evidence that both are useful in studying structure and 
evolution of a multivariate time series.

\end{abstract}

\begin{description}
\item[Keywords: ] Principal component analysis, heat map, FTSE 250, meteorological time series
\item[JEL Codes: C18] 
\end{description}

\section{Introduction}

Principal Components Analysis \citep{Jolliffe1986}  (PCA) is a multivariate
statistical technique used for dimension reduction
and to further an understanding of the underlying groups of variables in the data
(i.e. variables which may measure the same or similar underlying effect)
by extracting an ordered set of uncorrelated
sources of variation in a multivariate system. 

Often the objective in running a PCA is
to reduce the dimensionality of a data set while minimising the
loss of information. For example, if a dataset $\bf{x}$
has $p$ variables we would like to replace the $p$ variables
with $m$ principal components (PCs) where, ideally, $m \ll p$. A
critical question is then -- how small can we make $m$ without
an unacceptable loss of information?

Several standard methods exist. None of them are based on a statistical
test but instead they are ``rules of thumb'' for deciding on a suitable cutoff. 
One of the rules is based on the cumulative percentage explained, i.e.
retain the components which capture, say, 70\% or 90\% of the variation. 
Similarly, Kaiser's rule \citep{kaiser1960} is based on the principal of 
retaining components 
which have greater than or equal power to explain the data than a single variable.
Two other methods, the scree plot \citep{cattell1966} and log scree plot
\citep{farmer1971}, are based on looking for 
a change in behaviour in the plot of the variance explained (or its log).

Often, after the number of PCs to retain has been determined,
they are examined by a subject matter
specialist to determine if the PCs have an identifiable meaning. 
In the context of multivariate time series analysis a second
question arises after having made the selection, that is 
-- do the PCs have the same meaning
through out the entire sample period? 
In this paper we propose a two new methods aimed at simultaneously
selecting the number
PCs to retain and determining if their meaning has changed over time.

In recent years PCA has been widely applied
to the study of financial markets and we have chosen to highlight this 
method with the first  example from this field. 
Given that financial
markets are typically characterised by a high degree of multicollinearity,
implying that there are only a few independent sources of information
in a market, the uncorrelated nature of the eigenvectors
extracted by PCA make it an attractive method to apply.

With regards to financial time series, using 
the spectral decomposition theorem \cite[][p13]{Jolliffe1986},
many authorities have divided the eigenvectors into three distinct
groups based on their eigenvalues. For example,
\cite{kim2005} decomposed a correlation matrix of 135 stocks
which traded on the New York Stock Exchange (NYSE) into three parts:
\begin{enumerate}
\item The first principal component (PC1) with the largest 
eigenvalue which they asserted represented
a market wide effect that influences all stocks.
\item A variable number of principal components (PCs) following the 
market component 
which represented synchronised fluctuations affecting 
groups of stocks.
\item The remaining PCs indicated randomness in the price fluctuations.
\end{enumerate}
Therefore the questions are: how many components reflect 
the market and group components and, further, particularly for the
groups of stocks, do the PCs have the same meaning
through out the entire sample period? 

Our second example looks daily maximum temperatures from 
105 meteorlogical stations in Australia from 1 March 1975 to 31 December 2015. 
The stations
were chosen based on the completeness of the records. 

Similar to financial markets we expect the first principal component 
to represent overall weather conditions noting that we expect the maximum
daily temperatures to have a strong seasonal component. From there we are 
interested in how many principal component present meaningful 
components.

This paper adds to the literature by 
presenting, in addition to  some of the most commonly used methods for 
selecting the number of components to retain for
further analysis, 
two additional methods, one using heatmaps and the other a
change in eigenvector angle, to understand the structure
of the data and assist in determining how many components to retain.

The remainder of the paper is structured as follows; Section (\ref{sec:Data})
describes the data used, Section (\ref{sec:Method}) describes
our proposed new methods, Section (\ref{sec:Results}) presents our results
and Section (\ref{sec:Discussion}) contains our discussion and conclusions.

\section{Data}\label{sec:Data}

In this paper we present analysis based on two data sets, one financial and one meteorological. 

The London Stock Exchange FTSE 250 Index is a capitalisation-weighted index. 
It consists of the 101st to the 350th largest listed companies. 
Our data set covers the stocks in the FTSE 250 index as at 30 July 2015, and prices were sourced for the period 28 July 2000 and 30 July 2015 inclusive,
a fifteen year period with 3914 trading days. We downloaded the data
from Datastream and converted these data to a return series for
each stock. We identified which stocks had a complete history 
for the fifteen year period,  and retained 
a total of 147 stocks. The remaining stocks were  listed after 28 July 2000.
No allowance was made for the payment of dividends.

Our second example looks at daily maximum temperatures across the Australian
continent. 
Over time there have been 1877 stations  maintained by the
Australian Buerau of Meteorology. 
A complete list of meteorlogical sites was be obtained from the
Bureau's web site and 27 Sept, 2016.
The oldest station is the Melbourne Regional Office with records dating from
May, 1855. 
The completeness of their records range from 1\% to 100\%. 

Our data is the daily maximum
temperatures from 105 stations in Australia from 1 March 1975
to 31 December 2015. The stations
were chosen based on the completeness of the records over the period of 
investigation. 
The furthest East is Cape Moreton Lighthouse, Queensland, 
the furthest West is Carnarvon Airport, Western Australia,
 while the furtherest North is
Darwin Airport, Northern Territory and the furtherest South is Cape Bruny
Lighthouse, Tasmania.

\section{Methods}\label{sec:Method}

In multivariate time series analysis a PCA can be used to generate insights
into the key components of the variable set. 
One of the key considerations in applying such a tool is how many principal components
should be retained for further analysis. Below we describe three key steps in using heat maps 
to analyse how many components should be retained. We also briefly describe the 
standard methods used for this type of analysis.

\subsection{Construction of the Heat Maps}

In this paper we use a heat map to profile the change in 
the principal component coefficients over time. 
The visual nature of the output allows an examination of the heatmap
for consistency and patterns.

The three key steps in constructing the heat map are:
\begin{enumerate}
\item Determine an appropriate window size;
\item Calculate the coefficients using a sliding window; and
\item Present the heat map.
\end{enumerate}

Each of these steps is discussed below.

\subsubsection{Deciding on an Appropriate Window Size}\label{sub:Window}

In this study the window size is the minimum number of days (for the
meteorological application) or trading days (for the financial application)
for which the Kaiser-Meyer-Olkin \citep{kaiser1970,kaiser1974} measure of 
sampling adequacy (KMO) does not fall below 0.5. 
The KMO value is calculated as 
$$
\text{KMO}=\frac{\sum\sum_{j\ne k}r_{jk}^2}{\sum\sum_{j\ne k}r_{jk}^2+\sum\sum_{j\ne k}
q_{jk}^2}
$$
where the $r_{jk}$ are the original off-diagonal correlations and
the $q_{jk}$ are the off-diagonal elements of the partial-correlation
matrix. Thus the KMO statistic is a measure of how small the partial 
correlations are, relative to the original  correlations, the smaller
$\sum\sum_{j\ne k} q_{jk}^2$ are, the closer the KMO statistic will be to one.

A KMO value of $0.5$
is the smallest KMO value that is considered acceptable for a PCA.
The KMO test was performed using functions in the {\tt R} package
{\tt psych} \citep{psych}.

We have applied the KMO measure to both of our examples.

\subsubsection{Calculating the Coefficients}\label{sub:Coefficients}

PCA can be applied to either a correlation matrix or a covariance matrix.
All PCAs reported in this paper were carried out on correlation matrices.
PCAs were carried out using standard functions in
{\tt R} \citep{R}. For a correlation
matrix the total variation is equal to the number of variables
in the matrix. Correlation matrices were generated from the return series with the {\tt cor}
function in the {\tt stat} package in base {\tt R}.

The individual coefficients of the stocks or stations within each
component are subject to two mathematical constraints
by the PCA. The first is if $\alpha_i$
is the coefficient of the stock  or temperature station
$i$ then $-1\le \alpha_i \le 1$
and the second that
$$
\sum_{i=1}^N\alpha_i^2=1
$$
where $N$ is the number of stocks or stations in the sample. 

Correlation matrices were calculated for the first $k$ observations 
where $k$ was determined by the window size. Then a sliding
window approach is used to calculate the coefficients for each 
set of observations. With stock data this means that the first set of
coefficients is based on the first $k$ trading days and the next set is 
base on the second trading day to $k$ + 1 days and so on until the 
end of the period of investigation. With our daily maximum temperatures
the first $k$ observations are the first $k$ days since 1 March 1975.
The resulting matrix has the 
number of rows equal to the number of stocks or stations and the number of 
columns is the number of trading days or days less the window size. The 
matrix contains  the coefficients. 
It is these matrices that are visualised using a heat map. 

%

\subsubsection{Displaying the Heat Map}

To examine the time evolution of the PCs we
made heat maps  of  the $\alpha_i$'s 
using plotting functions in
the \verb+graphics+ package in base \verb+R+. The order of stocks
produced by the PCA matches the order in the input correlation matrix.
This is unlikely to be the most useful ordering so the stocks were sorted
and their order 
was fixed within each heatmap. In the  heatmaps of
the $\alpha_i$'s for stocks this was done by sorting on the coefficients at
the mid-point of the study period, but this choice of sorting point is
completely arbitrary. For the weather stations the ordering was based on the station number. 

\subsection{Variation of Eigenvector Direction}


The loadings of each Principal Component  are the 
coefficients of the eigenvectors in the  high dimensional data
space and so represent  a direction within that space. For example,
for  PC1, this is the direction of largest variation within the space.
Thus as an  addition or alternative
to the heatmaps, a second
method of examining the stability of the component meaning over time
is to calculate the change in direction of the eigenvector
in each subperiod. If this method is being performed instead of, rather than
in addition to,
generating a heatmap, one must first decide on the window size
and calculate the coefficients as described in Sections (\ref{sub:Window})
and (\ref{sub:Coefficients}).

We calculate the angle between the eigenvector in the
first subperiod and in each of the subsequent sub-periods as the
data window is slid across the data. The angle between two eigenvectors 
can be found from 

\begin{equation}
\theta=\arccos{({\bf a}\cdot{\bf b})}\label{eqn:Angle}
\end{equation}
The scalar product ${\bf a}\cdot{\bf b}$ would normally need to be
standardised by dividing by $|{\bf a}||{\bf b}|$ but the eigenvectors 
returned by the software we used were unit vectors rendering the
standardisation unnecessary. This generates a time series of angles
between the initial and subsequent eigenvectors which are plotted for
visual examination.

\subsection{Standard methods}

There are several standard methods used to evaluate how many 
principal components should be retained for further analysis. 
 Practitioners typically use more than one method 
and the methods often report a wide range of possible number of components to retain.

This section briefly  describes four methods; cumulative variance, 
Kaiser's rule,
scree plots, and log eigenvalue diagrams following \cite{Jolliffe1986} Ch.\ 6. 

The cumulative percentage of total variance is the percentage of variance explained 
by the first $m$ components. 
If $l_k$ is the variance of the $k^{\text{th}}$ PC then
$m$ is chosen to be the smallest $m$ such that
$$
t_m=100\sum_{k=1}^ml_k\Bigg/\sum_{k=1}^Nl_k
$$
is greater than a preselected cut-off $t^{*}$. Typically $t^{*}$ is between
70\% and 90\%. If the PCA is performed on a correlation matrix this simplifies
to 
$$t_m=\frac{100}{N}\sum_{k=1}^ml_k.
$$
The cutoff is  a preset value rather than one determined via a statistical 
test. 

The number of components this approach retains varies with the application 
but the authors' experience with financial time series data is 
that this method typically retains many components.


If the PCA is performed on a correlation matrix, then Kaiser's rule
\citep{kaiser1960} can be applied. Kaisers rule selects all PCs for which
$l_k>1$. This is based on the fact that each PC for which $l_k\le 1$
explains the same or less variation than a single variable on its
own. 
Variations of Kaiser's rule do exist for covariance matrices,
common ones are to select all PCs for which $l_k>\bar{l}$ or sometimes
more conservatively $l_k>0.7\bar{l}$, where $\bar{l}$ is the mean
value of the $l_k$.

\cite{cattell1966} discussed using a scree graph
which has the component number of the x-axis and the variance associated
with each component on the y-axis, plotted as a line graph. 
That is, it is a plot of
$l_k$ against $k$. 
The practitioner then 
looks at the graph to determine where the ``elbow'' in the graph is such that it is ``steep'' 
to the left and ``shallow'' to the right. For some applications this decision may be
subjective.  

A simple variation on this, discussed by \cite{farmer1971}, is to plot 
$\log l_k$, instead of $l_k$. 
That is the log-eigenvalue diagram
which plots  the component number of the x-axis and the log of the variance associated
with each component on the y-axis, plotted as a line graph. 
As with the scree plot, the practitioner then 
looks at the graph to determine where cut off should be, this time
looking for the point at which the decay becomes linear.


\section{Results}\label{sec:Results}



This section presents the results of the heat maps and
angle change analysis and the outcome
of deciding on the number of components using scree plots, log value diagrams,
cumulative percentage, Kaiser's rule. All PCAs reported 
in this paper were carried out on correlation matrices.

\subsection{Heat Maps}\label{sub:Heat}

As discussed in Section (\ref{sec:Method}) the three key steps in 
constructing the heat map are: determine 
an appropriately sized window using the KMO statistic; 
calculate the coefficients; and present the heat map. 

For the stock market application if the window size was 
200 trading days then KMO value could not be estimated for some windows.
If the window size was 250 then minimum the KMO value was 0.69. 
Therefore for this application we used a window size of 250 trading days. 

For the maximum temperature application if the window size was 
eight years (2920 days) then the minimum the KMO value was 0.99.
With fewer years of data the KMO statistic could not be estimated for
some windows.  
Therefore for this application we used a window size of 8 years or 2920 days.

The coefficients correlation matrices 
were calculated 
 from the return series 
on a rolling window of 250 trading days for the stock application
(giving 3664 sets of
coefficients, with each set having 147 individual coefficients)
and the rolling window of 2920 days for the temperature application (giving 
11996 sets of coefficients with 105 individual coefficients).

\begin{figure}[ht]
  \centering
  \includegraphics[width=14cm]{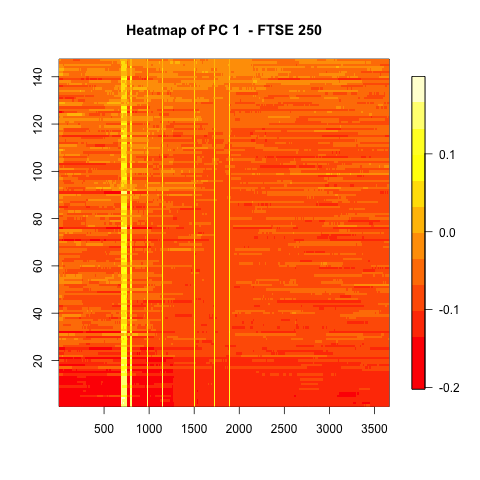}
  \caption{Heatmap of coefficients of PC 1 for the financial application.
The horizontal axis is in trading days. The vertical axis is the 147 stocks
in the sample sorted by the value of their coefficients in the PCA at the
mid-point of the sample.}
\label{fig:HeatPC1}
\end{figure}

\begin{figure}[ht]
  \centering
  \includegraphics[width=14cm]{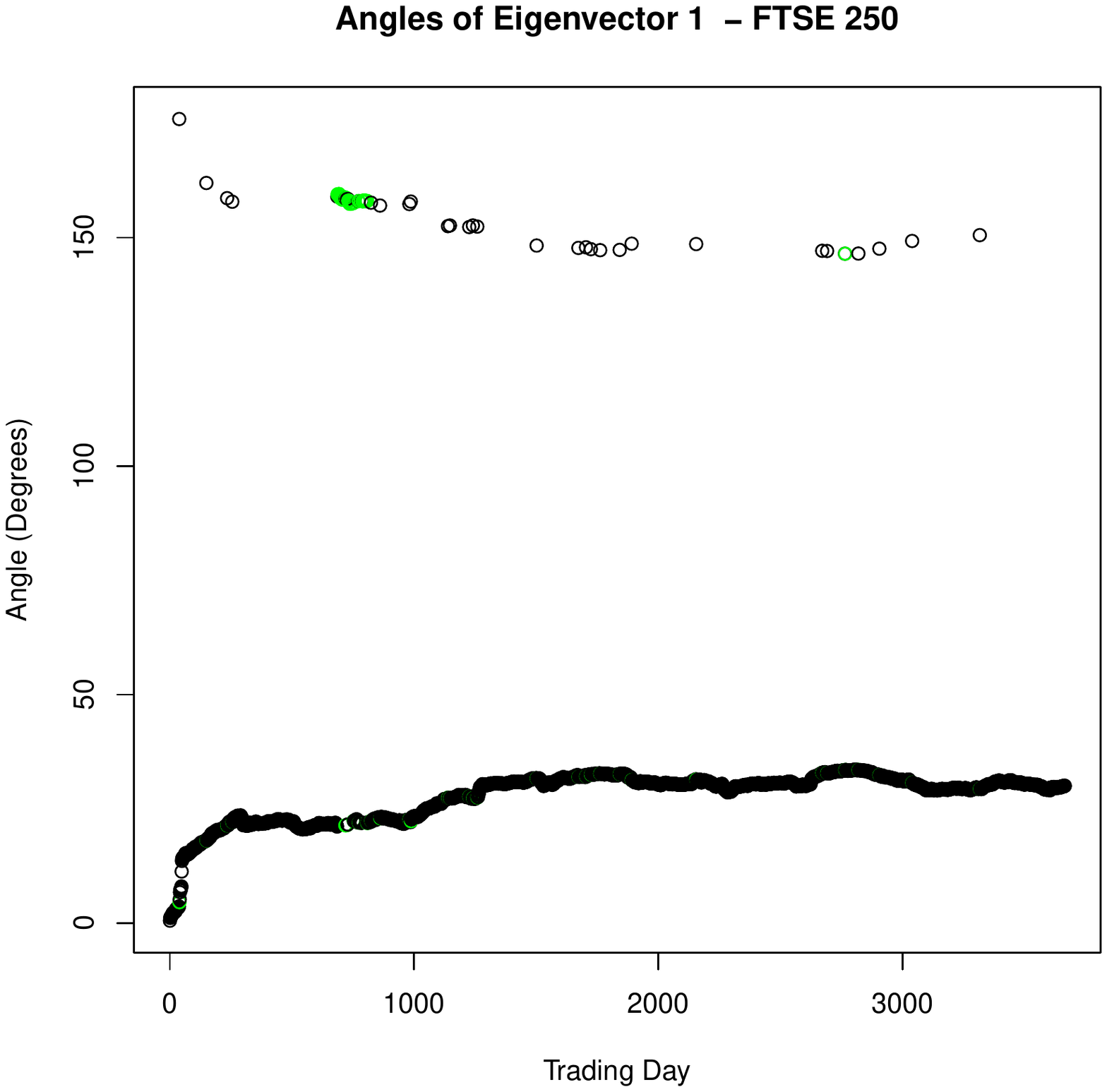}
  \caption{Angles of eigenvectors of PC 1 between the first and
subsequent periods. The points are colour coded black indicates
a loading of less than or equal to zero on stock 147 in the 
heatmap in Figure (\ref{fig:HeatPC1}) while green indicates
a loading of greater than zero.
 }
\label{fig:AnglesPC1}
\end{figure}

\begin{figure}[ht]
  \centering
  \includegraphics[width=14cm]{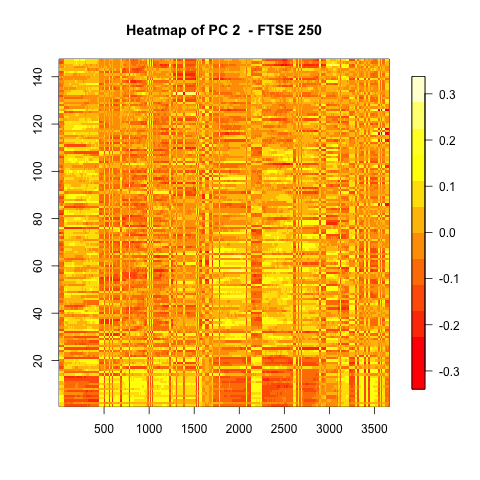}
  \caption{Heatmap of coefficients of PC 2.
 }
\label{fig:HeatPC2}
\end{figure}

\begin{figure}[ht]
  \centering
  \includegraphics[width=14cm]{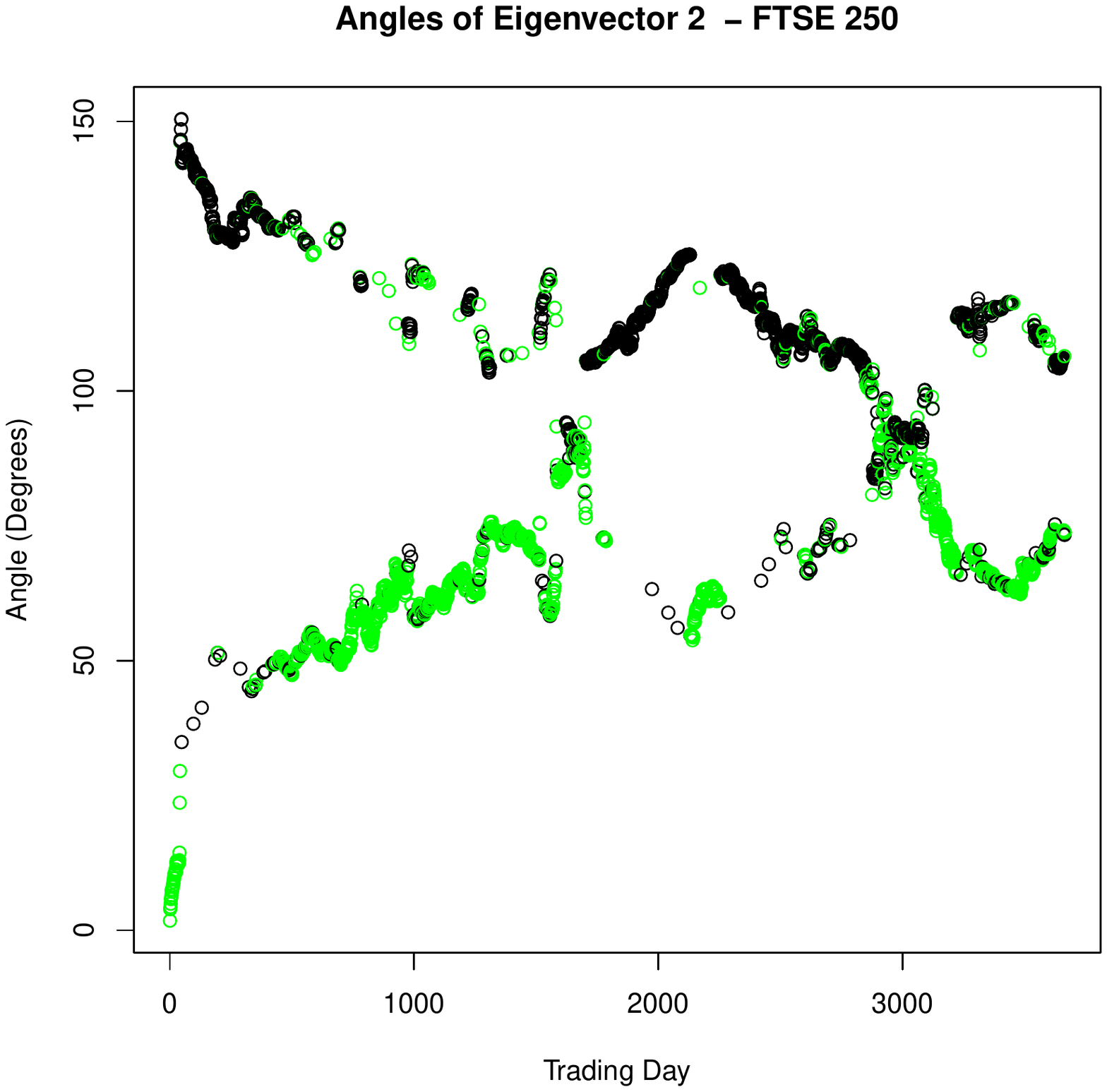}
  \caption{Angles of eigenvectors of PC 2  between the first and
subsequent periods. The points are colour coded black indicates
a loading of less than or equal to zero on stock 147 in the
heatmap in Figure (\ref{fig:HeatPC2}) while green indicates
a loading of greater than zero. 
 }
\label{fig:AnglesPC2}
\end{figure}

To examine the time evolution of the principal components we
present heat maps  of  the $\alpha_i$'s. 
Each heat map has time on the x-axis because each set of coefficients was
obtained
from the PCA applied to the rolling window. The y-axis has the 147 stocks 
for the stock application or the 105 weather stations for the temperature 
application. 
For the stock application these were
 sorted by the coefficients at the mid-point of the study period for 
heatmaps of the $\alpha_i$'s
and for temperature it was sorted by station ID. 
Note that we call the coefficients structure with 
lower coefficients at the top  ``structure A'' and 
the higher coefficients at the top the ``structure B''.

The key focus of this subsection is on using heat maps to determine the 
number of components to retain for further analysis. Below we focus 
in detail on each of the applications. 

For the stock market application clearly the first 
component (Figure \ref{fig:HeatPC1})
has a structure and offers useful insights into the underlying data.
Above we noted that in financial applications the first principal component 
is considered by other authorities to be the market wide 
effect. 
An examination of principal component one showed the majority of
the heat map represents structure B as the reds (negative coefficients) are at the bottom.
This heat map is  consistent with the hypothesis that the
first component captures a single market-wide effect. 

For the stock market application beyond the first component
the heat maps show the coefficients 
switch between structure A 
and structure B and this switching happens increasingly 
rapidly for the higher numbered components. For reasons of space we
only present the heat map for PC2 in Figure (\ref{fig:HeatPC2}), the
remainder are available on request from the authors.
The heat map for PC2 is a mixture of  structure A with blocks
of structure B. The heat map for principal component three had some
small sections which are dominated by structure A  however some blocks are dominated by
rapid change between structure A and structure B. 
The heat map for principal component four is similar to that of principal component three.
The heat map for principal component five is dominated by rapid change with a few other 
blocks of consistent behaviour, as is component six. The components beyond 
component six were dominated by rapid change between structure A and B.
 
Therefore, based on this analysis the maximum number of components to retain 
for the stock market application would be at most
six. However, it would also be reasonable to keep just PC1 noting that 
only the first component could reasonably be considered to have a nearly 
constant fundamental structure over the sample period.

\begin{figure}[ht]
  \centering
  \includegraphics[width=14cm]{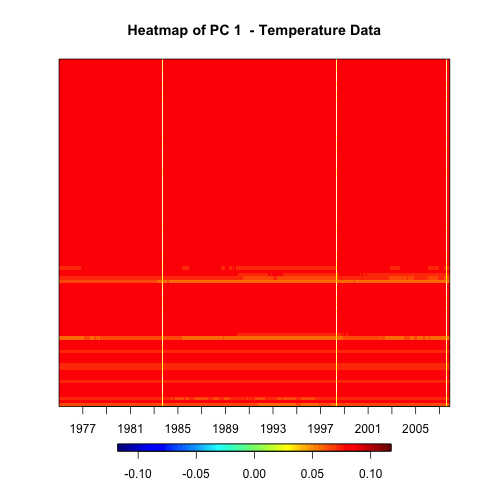}
  \caption{Heatmap of coefficients of PC 1.
 }
\label{fig:HeatBOM_PC1}
\end{figure}

\begin{figure}[ht]
  \centering
  \includegraphics[width=14cm]{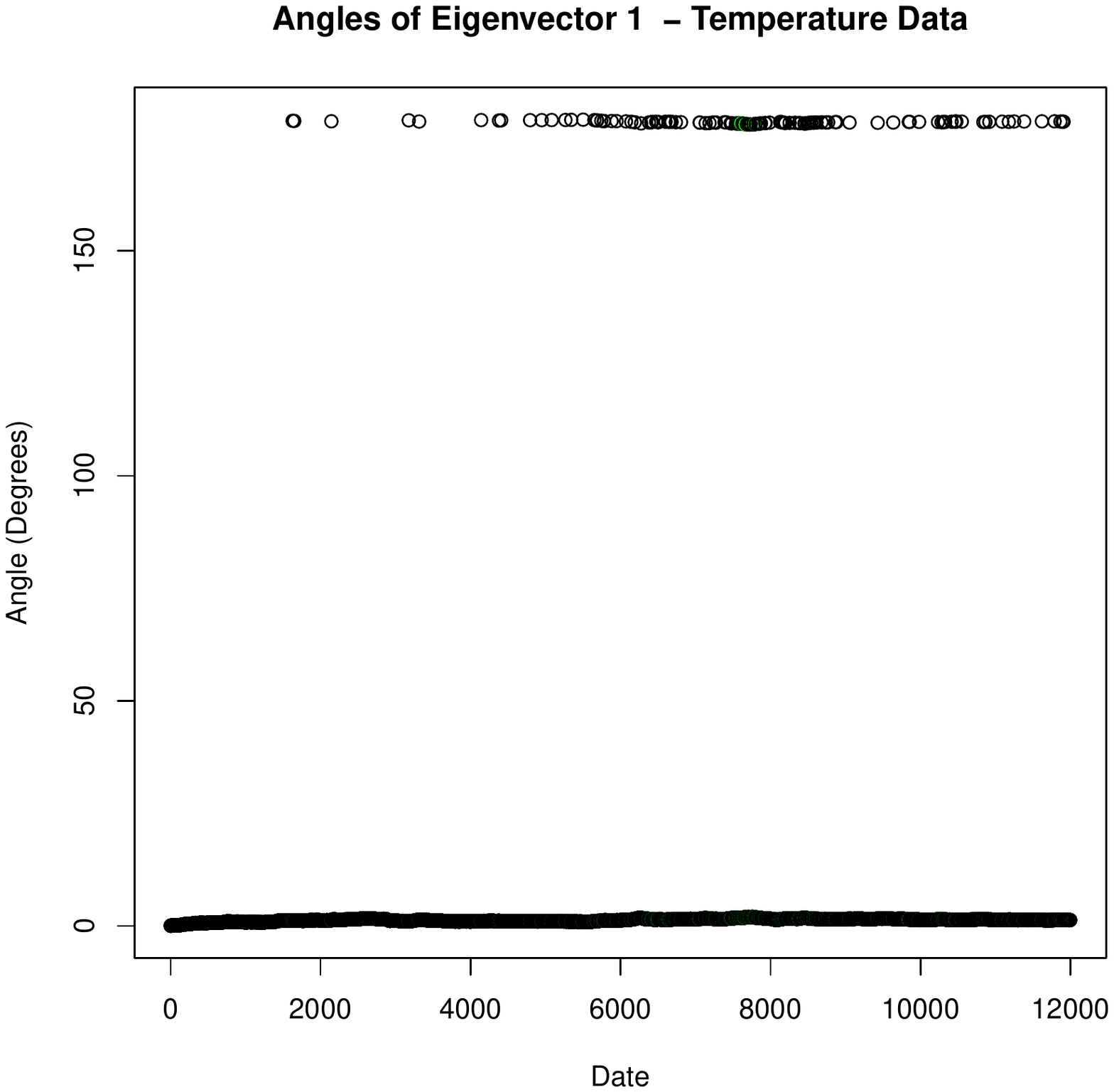}
  \caption{Angles of eigenvectors of PC 1.
 }
\label{fig:AnglesBOM_PC1}
\end{figure}

\begin{figure}[ht]
  \centering
  \includegraphics[width=14cm]{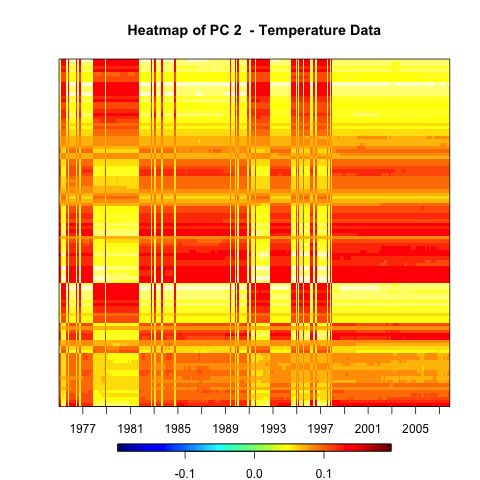}
  \caption{Heatmap of coefficients of PC 2.
 }
\label{fig:HeatBOM_PC2}
\end{figure}

\begin{figure}[ht]
  \centering
  \includegraphics[width=14cm]{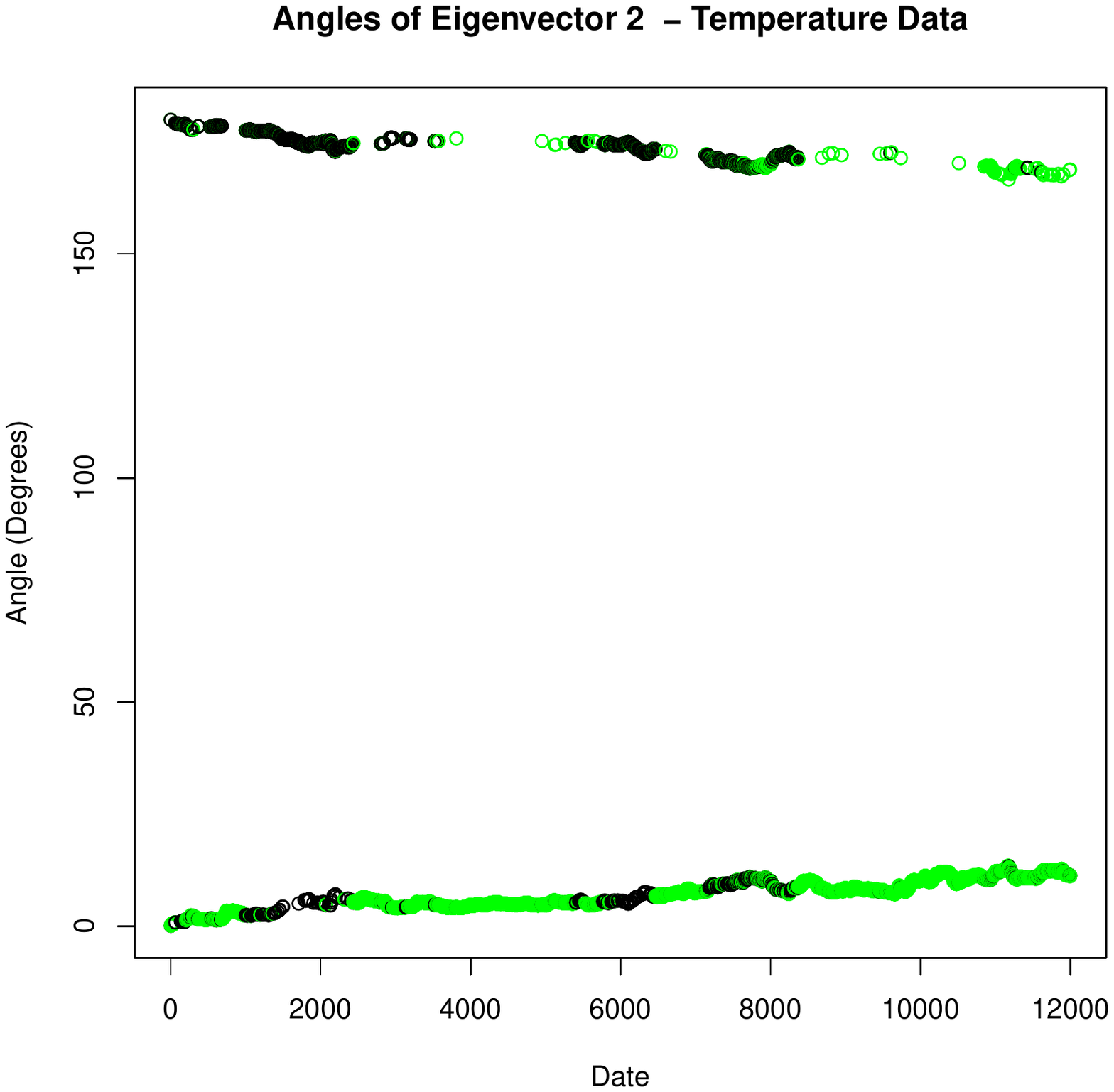}
  \caption{Angles eigenvectors of PC 2.
 }
\label{fig:AnglesBOM_PC2}
\end{figure}

\begin{figure}[ht]
  \centering
  \includegraphics[width=14cm]{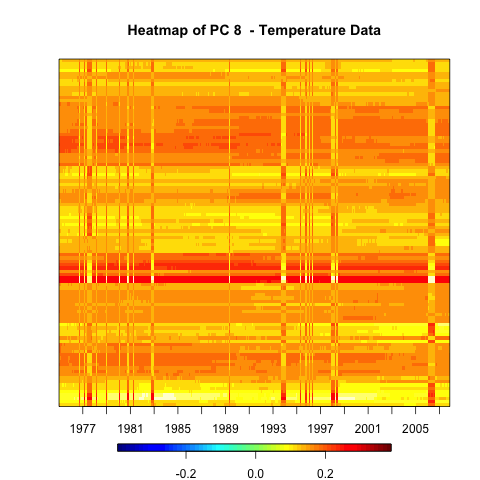}
  \caption{Heatmap of coefficients of PC 2.
 }
\label{fig:HeatBOM_PC8}
\end{figure}

\begin{figure}[ht]
  \centering
  \includegraphics[width=14cm]{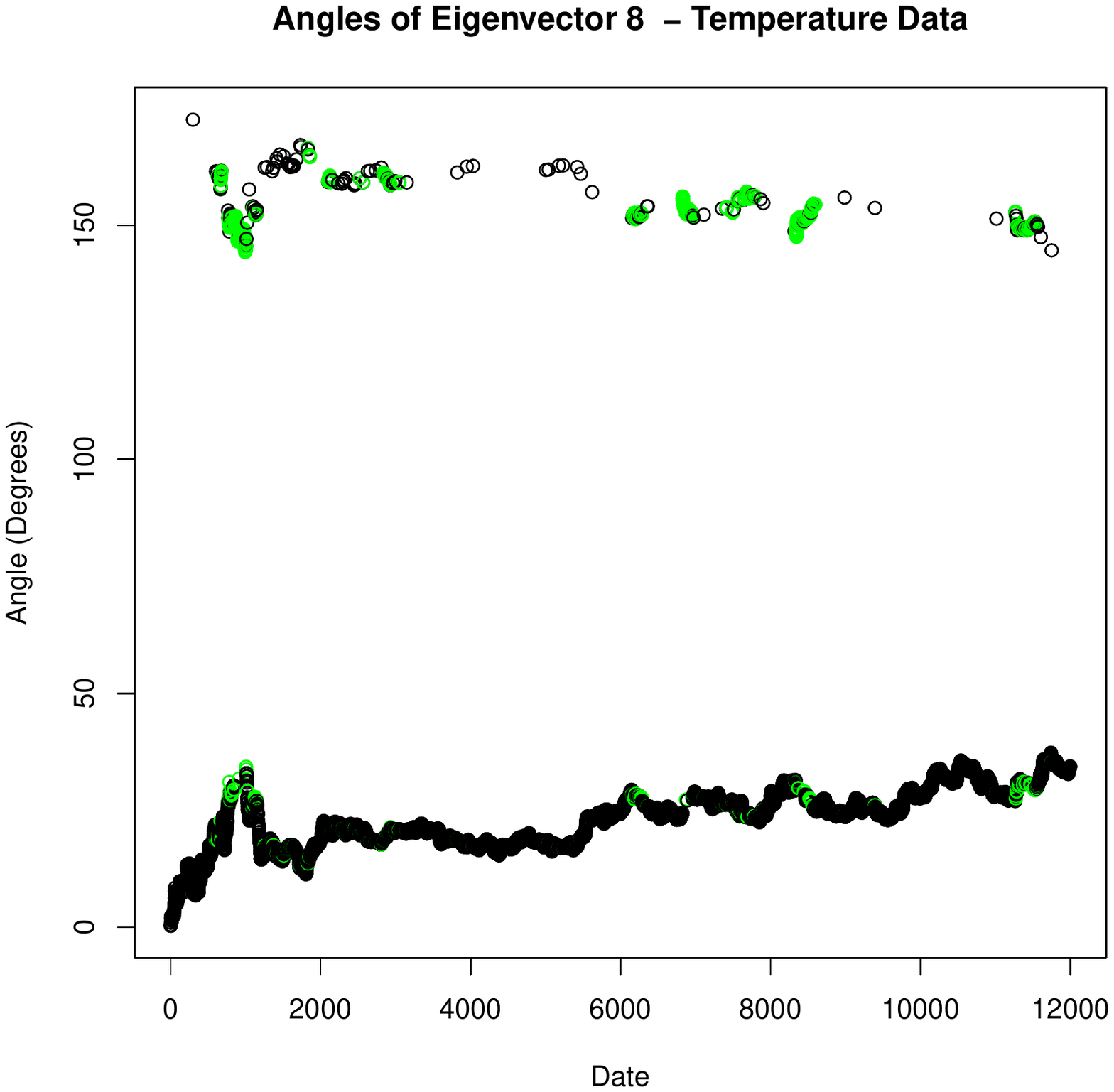}
  \caption{Angles eigenvectors of PC 8.
 }
\label{fig:AnglesBOM_PC8}
\end{figure}

\begin{figure}[ht]
  \centering
  \includegraphics[width=14cm]{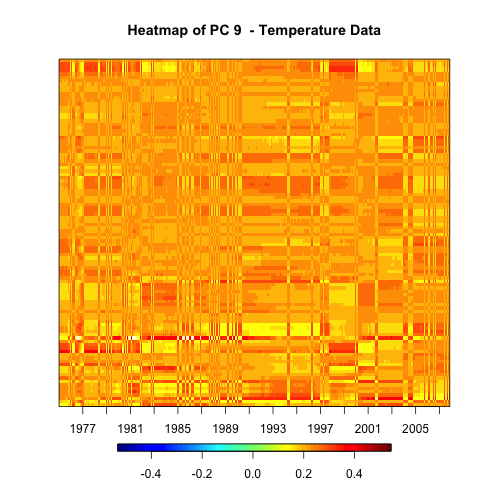}
  \caption{Heatmap of coefficients of PC 9.
 }
\label{fig:HeatBOM_PC9}
\end{figure}

\begin{figure}[ht]
  \centering
  \includegraphics[width=14cm]{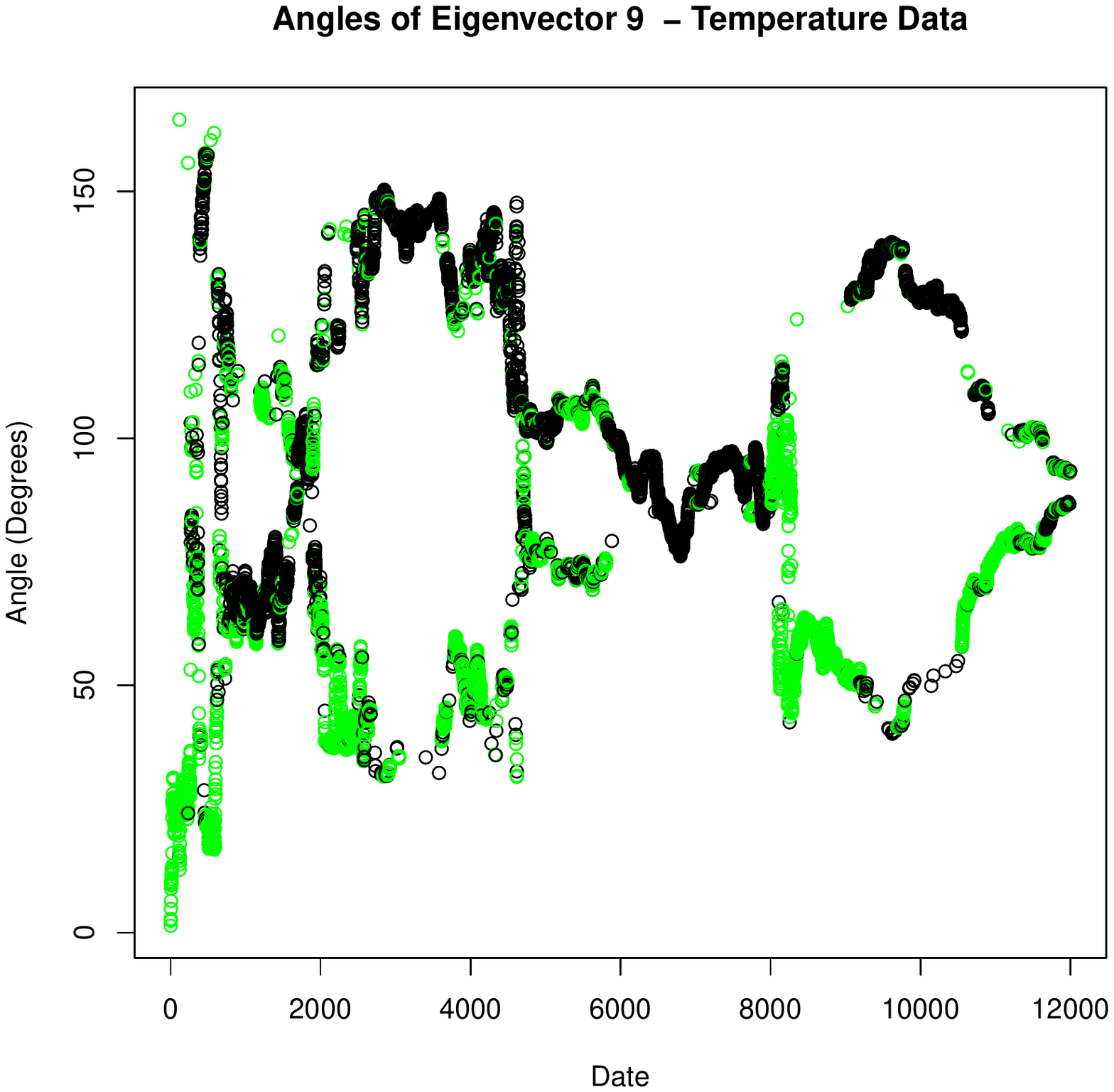}
  \caption{Angles eigenvectors of PC 9.
 }
\label{fig:AnglesBOM_PC9}
\end{figure}

For the temperature application 
we see a similar behaviour in the first component in
that the majority of the heatmap is red.
This is likely to represent the similarity in the daily maximum 
temperature despite the variation across the continent.

For the daily maximum temperature application 
beyond the first component
the heat maps show the coefficients 
switch between structure A and structure B.  The heat map for principal component two 
is dominated by structure A until the sliding window starting in about 1995 
and then  is dominated
by blocks of structure B. It is worthwhile noting that there is a set of weather stations
with the exact opposite pattern to this majority (which are mainly red until 1995 
and mainly yellow thereafter).

The heat map for principal component three has a section from 
the sliding window beginning in 1990 to about 1996 which is different from 
the majority of the rest of the component. In this section there is rapid change
between structure A and structure B. 

The heat map for principal component four is  dominated by structure B
especially  from the sliding window beginning in 1989, although 
the stations with higher numbers do have the lower coefficients. 

The heat map for principal component 5 is dominated by structure A
with a noticeable section of different behaviour in the sliding windows beginning
in 1993 to 2001. Interestingly this component also contains some time invariant 
behaviour with clear blocks of stations.

The heat map for principal component 6 has some sections of  rapid change 
with a consistent period for the sliding windows beginning in the 1980's. From 
here the components begin to have increasingly rapid changes. 

\subsection{Change in Eigenvector Angle}

The details of the selection of the window size and calculation
of coefficients for the calculation of the eigenvector angles
is identical to that described in Section (\ref{sub:Heat}) above.

The results for the FTSE 250 index example are presented in
Figures (\ref{fig:AnglesPC1}) and (\ref{fig:AnglesPC2}). The 
results for the temperature data are presented in 
Figures (\ref{fig:AnglesBOM_PC1}), (\ref{fig:AnglesBOM_PC2}), 
(\ref{fig:AnglesBOM_PC8}) and (\ref{fig:AnglesBOM_PC9}).
For reasons of space we have not included the results for
PCs three through seven for the temperature data
but these are available on request.

For the FTSE 250 data, the angle change for PC1 presents a picture that
is consistent with the heat map in Figure (\ref{fig:HeatPC1}) in that
after an initial change in the angle of the eigenvector in the
earliest period in the data it remained stable over the remainder
confirming the observation that PC1 is likely to have a single meaning
over the whole of the sample period. The angle change for PC2 
(Figure \ref{fig:AnglesPC2})
presents a picture which supplements the heat map in Figure (\ref{fig:HeatPC2})
showing that after the initial rapid change in angle at the beginning
of the sample period the angle continues to change over the remainder
of the period. The two structures seen in Figure (\ref{fig:HeatPC2}) seem
to reflect two directions within the data space, but even within
that structure the direction was
changing over time. Given the change in angle only PC1 can be considered
to have retained its meaning over the entire sample period.

For the temperature data, the plots of the angles of the eigenvectors 
show that the direction of the variation with the data space changes
slowly for PCs one through eight. With PC9 there is a substantial
change in the behaviour with large changes in angle over time.
From this we can consider PCs one through eight to
have the same meaning over the sample period.

\subsection{Standard methods}

We now turn to comparing these results to traditional methods.

We applied cumulative variance with four thresholds present
the outcomes for the number of 
principal components to retain for further analysis in Table \ref{tab:cumvar}.

\begin{table}[h]
\centering
\caption{Results of applying cumulative variance with four different thresholds}
\label{tab:cumvar}
\begin{tabular}{l|ll}
Threshold (\%) & Number of components  & Number of components \\ 
& Stock Market & Daily maximum temperature \\ \hline
60 & 22 & 0 \\
70 & 34  & 1\\
80 & 50  & 3\\
90 & 73  & 10
\end{tabular}
\end{table}

%
%
Applying Kaisers rule to the stock market application
with cutoff of 1 suggested that
38 components be retained for further analysis while with 0.7 it suggested 56.
For the maximum daily temperature application the number of components
to retain is 8 with a cutoff of 1 and 11 with a cutoff of 0.7. 

\begin{figure}[ht]
  \centering
  \includegraphics[width=7cm]{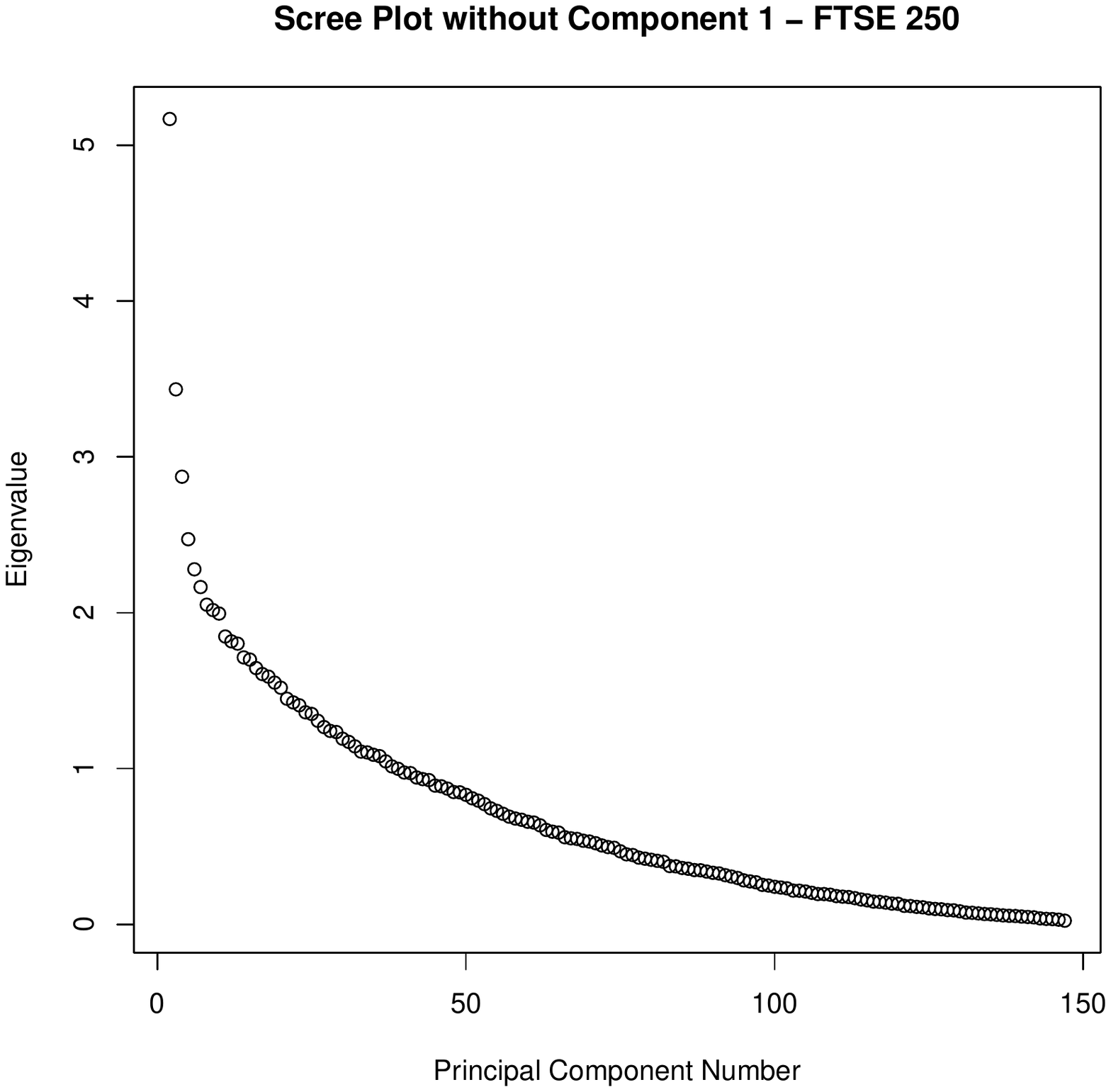}
  \includegraphics[width=7cm]{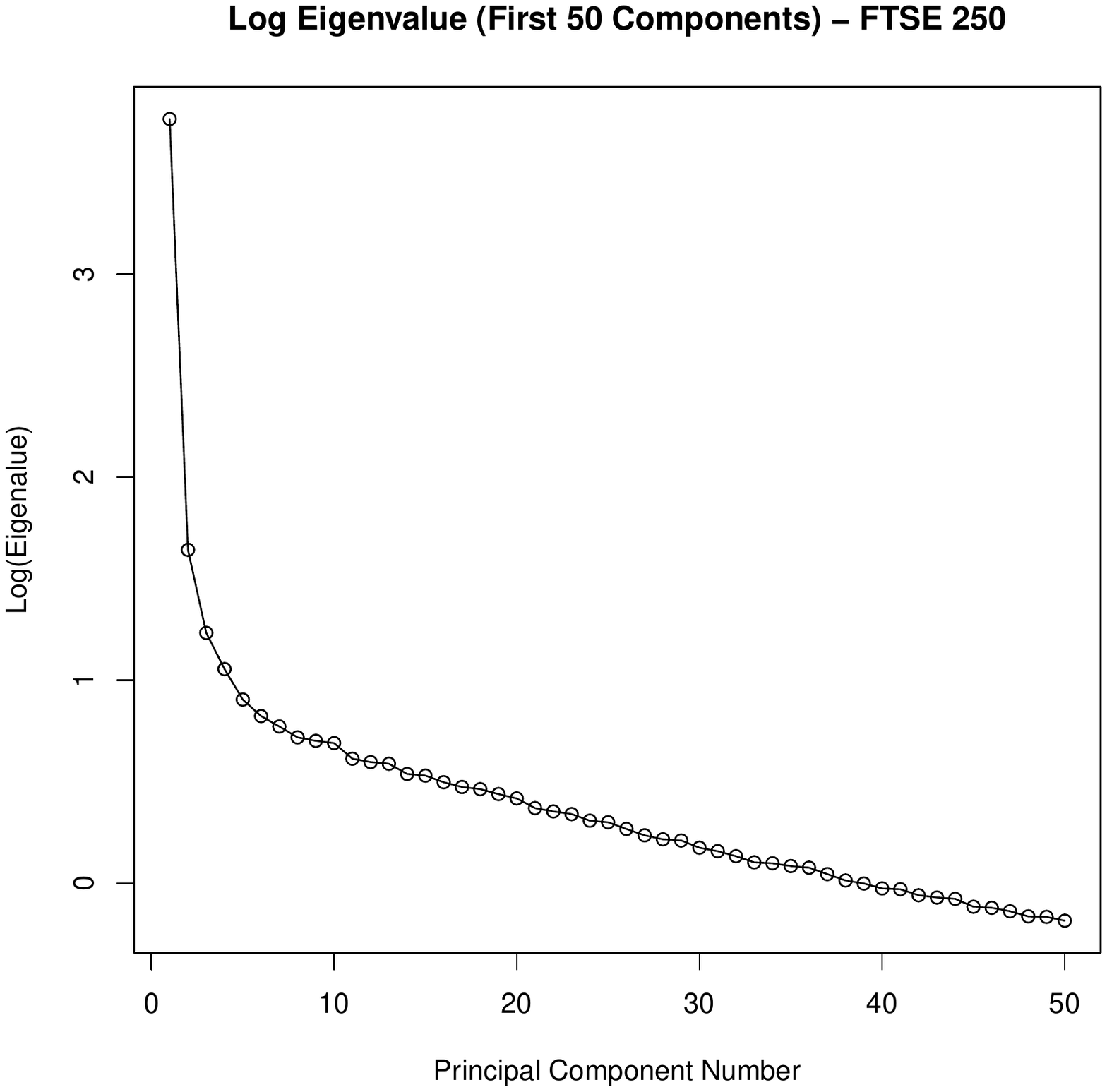}
  \caption{Scree plots for FTSE 250 data.
 }
\label{fig:ScreeFTSE250}
\end{figure}

\begin{figure}[ht]
  \centering
  \includegraphics[width=7cm]{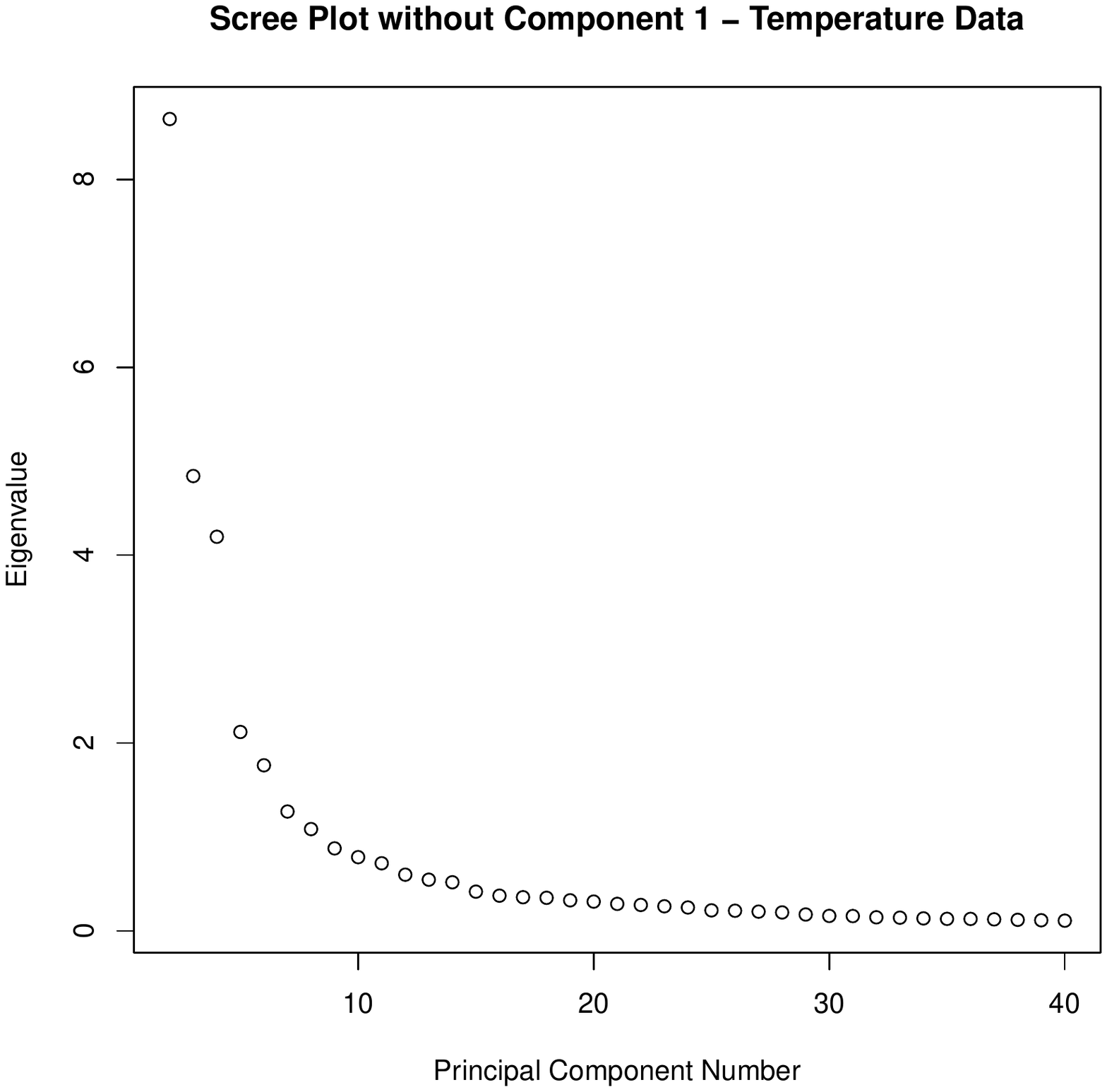}
  \includegraphics[width=7cm]{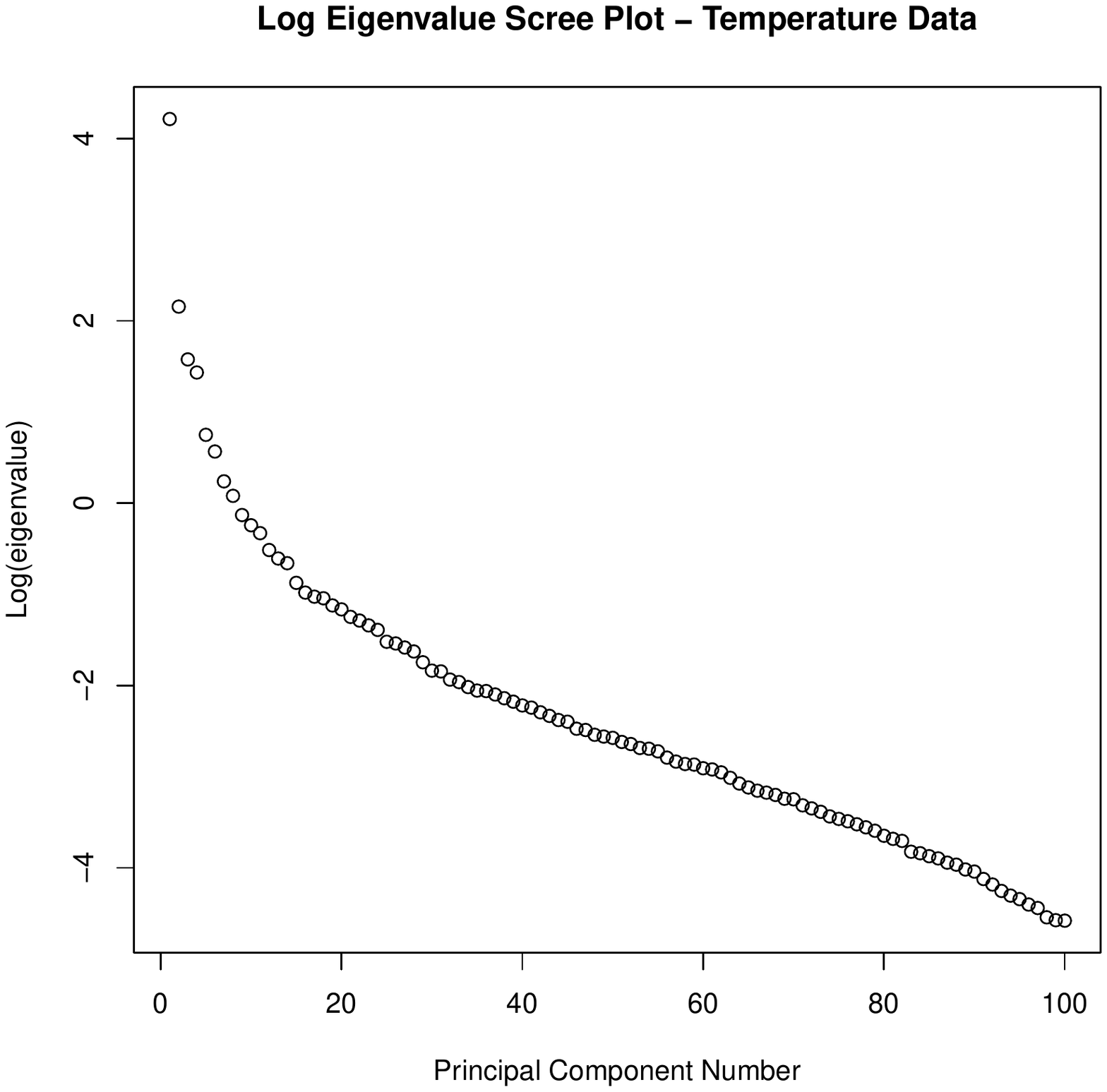}
  \caption{Scree plots for temperature data.
 }
\label{fig:ScreeBOM}
\end{figure}

The interpretation of the scree plots depends on the practitioner determining 
where the elbow is in the graph presented in the left panel of
Figures (\ref{fig:ScreeFTSE250}) and (\ref{fig:ScreeBOM}). For the 
stock market application (Figure \ref{fig:ScreeFTSE250}) one could 
suggest that there are breaks at either five, seven or 10. Retaining more than 10
components would
not be reasonable based on the scree plot. For the daily maximum temperatures 
application (Figure \ref{fig:ScreeBOM}) 
one could suggest that there are breaks at either two, four or six components. 

The interpretation of the log value diagrams depends on the practitioner determining 
where the decay becomes linear in the graphs presented in the
right panel of Figures (\ref{fig:ScreeFTSE250}) and (\ref{fig:ScreeBOM}). For the 
stock market application as with the 
scree plot, in this example one could 
suggest that there are breaks at either five, seven or 10. 
For the daily maximum temperature application the break could be at either one, 
two, four, six or eight components.


\section{Discussion}\label{sec:Discussion}

As noted above, when considering a PCA on time series data
it is important to question  whether the PCs have the same meaning
throughout the entire sample period. Subsequently we have 
presented two visualisation techniques; one of creating a heat map from a set
of eigenvalue coefficients created by applying a sliding window 
to the time series and a second by checking the change of the angle
of the eigenvector with respect to the first. 
The heat map can then be examined for evidence
of structure and the angle for change in direction along which the 
variation lies within the high-dimensional space. 

Our first example data set was from financial time series and comprised of 
3914 trading days of the FTSE 250 starting from 28 July 2000. Standard
methods of deciding the number of components varied highly from up to 73
for cumulative variance and 109 for Kaisers rule to down to 10 for scree plot and 
log plot. 
Our second example data set was from daily maximum temperature
series from 105 weather stations in Australia. For this analysis
we used data from 1 March 1975 to 31 December 2015. Standard
methods of deciding the number of components gave varying results
but not more than 10 components (this is primarily because the 
first component explains more than 60\% of the variation).  
Particularly for the stock market application 
these rules lead to very diverse recommendations and offer no insights
into the behaviour of each component.

Our suggestion is to compliment the standard methods with heat maps and
angle variation. The 
two methods 
we proposed each have three parts, the first
two in common: determine the appropriate window size, then
calculate the coefficients using a sliding window.
The heat map can be generated from the coefficients. We have found that
sorting variables order can improve the interpretability of the heat map.
The heat map offers valuable insights into the structure of each principal component
and when there is no longer a structure such principal components should not 
be retained for further analysis. 

As indicated in the introduction, a key question when performing a 
PCA on a multivariate time series is -- do the PCs have the same meaning
through out the entire sample period? This question is important both for
deciding how many components to retain from a PCA and understanding
their meaning. 

In the context of financial time series analysis
this is important because other authorities, such as \cite{kim2005},
do assign financial meaning to a number of components. 
The heat maps presented in Figures (\ref{fig:HeatPC1}) and (\ref{fig:HeatPC2})
can be usefully applied to answer
both questions. These figures
show that the first principal component
is dominated by structure A (lower coefficients) which is consistent with 
the hypothesis that the first component is a market-wide component. 
Components two
to six show some evidence of structure, 
though components beyond PC2 have increasing 
proportions of time where there is rapid change. 

While the results 
of the scree plot and log eigenvalue plot suggests retaining as many as 10
PCs, the heatmaps indicate that it is not really possible to assign
a financial meaning to any PCs beyond PC2. The angle variation analysis
suggests only PC1 cold have financial meaning. So while in a statistical
sense PCs one through 10 may be retained to explain variation in the
data set, a financial subject matter specialist would be unlikely to be
able to find useful information beyond one, or at most two, of these PCs.

Our second example applied the same methodology to 
daily maximum temperatures from 105 weather stations across Asutralia. 
These heat map for principal component one
was dominated by structure A (higher coefficients) and components two
through eight some evidence of structure, a conclusion supported by the
angle change analysis . These results were consistent with 
the standard methods, but once again the diagrams offered a greater 
level of detail.

Thus the heat maps of the PCs offers significantly more insights into 
the processes that generated the data over time than the other methods
commonly used to determine how many PCs to retain from a PCA. Further,
the angle change analysis also yielded useful insights.


\bibliography{references}

\begin{thebibliography}{}

\bibitem[\protect\citeauthoryear{Cattell}{Cattell}{1966}]{cattell1966}
Cattell, R.~B. (1966).
\newblock {The scree test for the number of factors}.
\newblock {\em Multivariate Behavioural Research\/}~{\em 1}, 245--276.

\bibitem[\protect\citeauthoryear{Farmer}{Farmer}{1971}]{farmer1971}
Farmer, S.~A. (1971).
\newblock {An investigation into the results of principal component analysis of
  data derived from random numbers }.
\newblock {\em Statistician\/}~{\em 20}, 63--72.

\bibitem[\protect\citeauthoryear{Jolliffe}{Jolliffe}{1986}]{Jolliffe1986}
Jolliffe, I.~T. (1986).
\newblock {\em {Principal Component Analysis}}.
\newblock New York: Springer.

\bibitem[\protect\citeauthoryear{Kaiser}{Kaiser}{1960}]{kaiser1960}
Kaiser, H.~F. (1960).
\newblock {The Application of Electronic Computers to Factor Analysis}.
\newblock {\em Educational and Psychological Measurement\/}~{\em 20}, 141--151.

\bibitem[\protect\citeauthoryear{Kaiser}{Kaiser}{1970}]{kaiser1970}
Kaiser, H.~F. (1970).
\newblock {A second generation little jiffy}.
\newblock {\em Psychometrika\/}~{\em 35\/}(4), 401--415.

\bibitem[\protect\citeauthoryear{Kaiser and Rice}{Kaiser and
  Rice}{1974}]{kaiser1974}
Kaiser, H.~F. and J.~Rice (1974).
\newblock {Little jiffy}.
\newblock {\em Educational and Psychological Measurement\/}~{\em 34\/}(1),
  111--117.

\bibitem[\protect\citeauthoryear{Kim and Jeong}{Kim and Jeong}{2005}]{kim2005}
Kim, D.-H. and H.~Jeong (2005).
\newblock {Systematic analysis of group identification in stock markets}.
\newblock {\em Physical Review E\/}~{\em 72}, 046133.

\bibitem[\protect\citeauthoryear{{R Core Team}}{{R Core Team}}{2014}]{R}
{R Core Team} (2014).
\newblock {\em R: A Language and Environment for Statistical Computing}.
\newblock Vienna, Austria: R Foundation for Statistical Computing.

\bibitem[\protect\citeauthoryear{Revelle}{Revelle}{2014}]{psych}
Revelle, W. (2014).
\newblock {\em psych: Procedures for Psychological, Psychometric, and
  Personality Research}.
\newblock Evanston, Illinois: Northwestern University.
\newblock R package version 1.4.5.

\end{thebibliography}

\newpage

\clearpage

\appendix


\section{Additional Heatmaps and Angles -- FTSE 250}

\begin{figure}[ht]
  \centering
  \includegraphics[width=14cm]{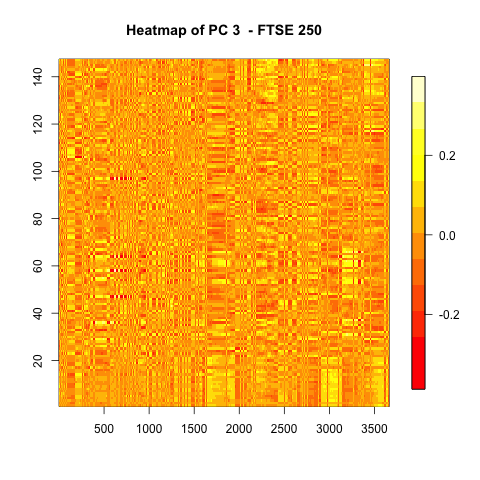}
  \caption{Heatmap of coeffients of PC 3.
 }
\label{fig:HeatFTSE_PC3}
\end{figure}

\begin{figure}[ht]
  \centering
  \includegraphics[width=14cm]{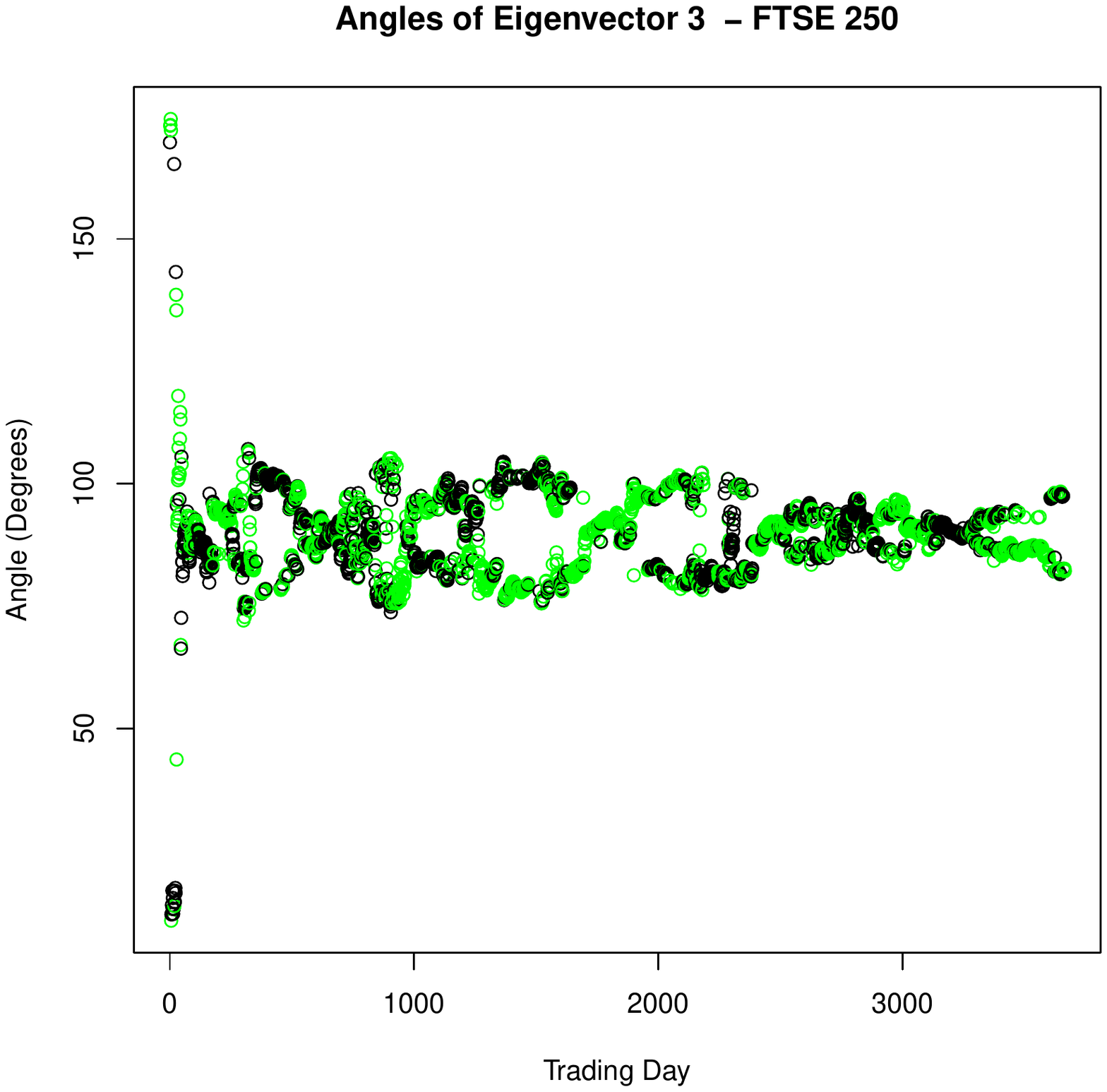}
  \caption{Angles of eigenvectors of PC 3.
 }
\label{fig:AnglesFTSE_PC3}
\end{figure}

\begin{figure}[ht]
  \centering
  \includegraphics[width=14cm]{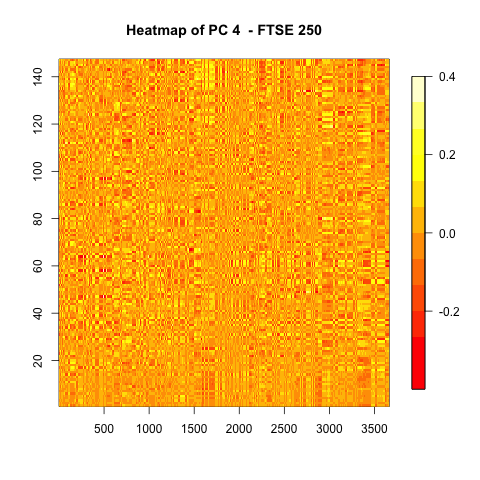}
  \caption{Heatmap of coeffients of PC 4.
 }
\label{fig:HeatFTSE_PC4}
\end{figure}

\begin{figure}[ht]
  \centering
  \includegraphics[width=14cm]{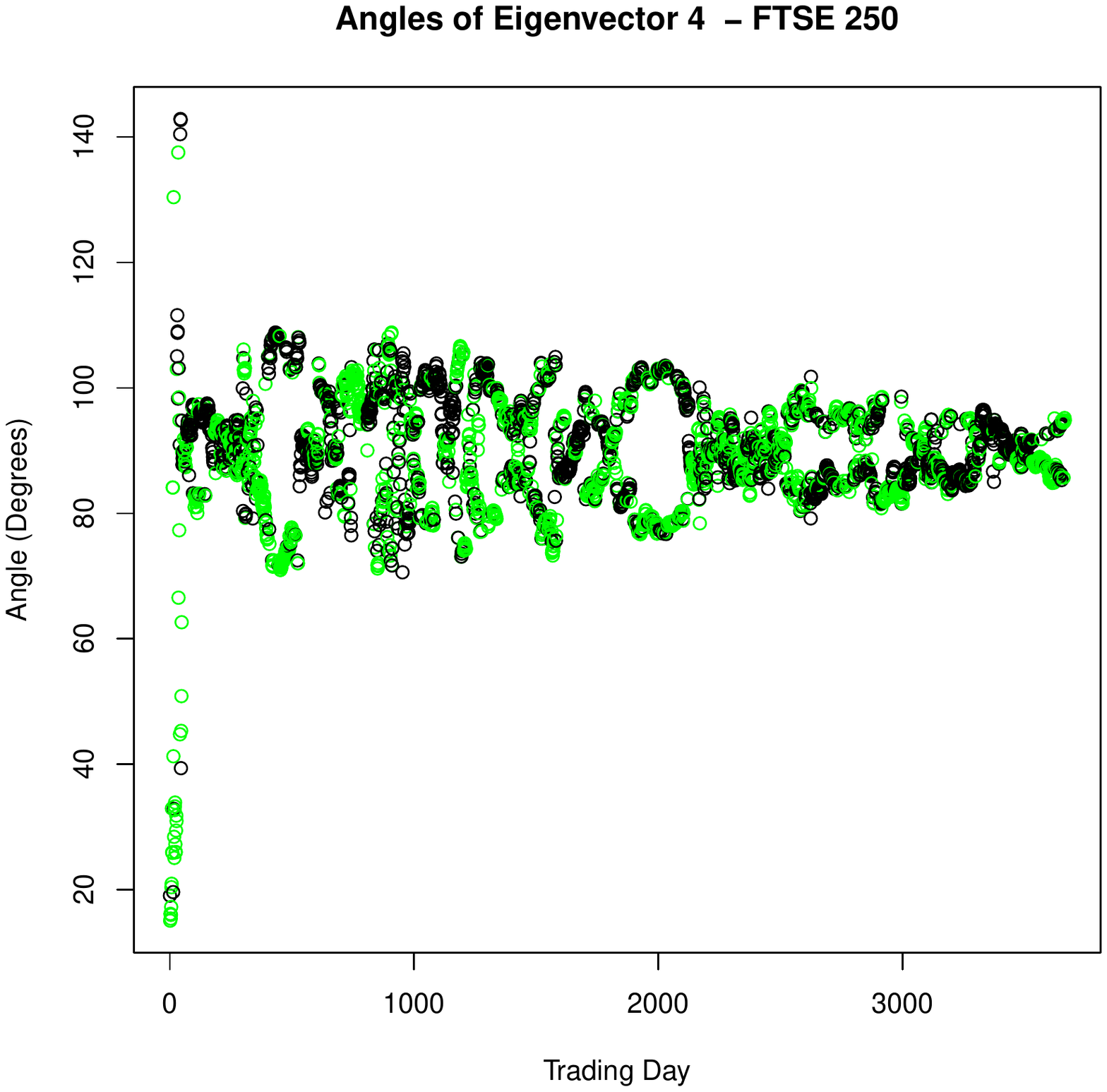}
  \caption{Angles of eigenvectors of PC 4.
 }
\label{fig:AnglesFTSE_PC4}
\end{figure}

\begin{figure}[ht]
  \centering
  \includegraphics[width=14cm]{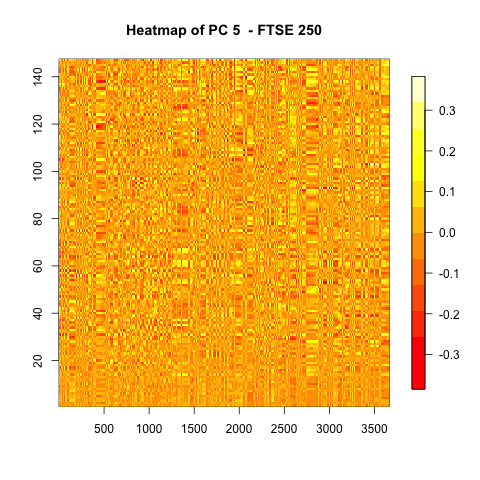}
  \caption{Heatmap of coeffients of PC 5.
 }
\label{fig:HeatFTSE_PC5}
\end{figure}

\begin{figure}[ht]
  \centering
  \includegraphics[width=14cm]{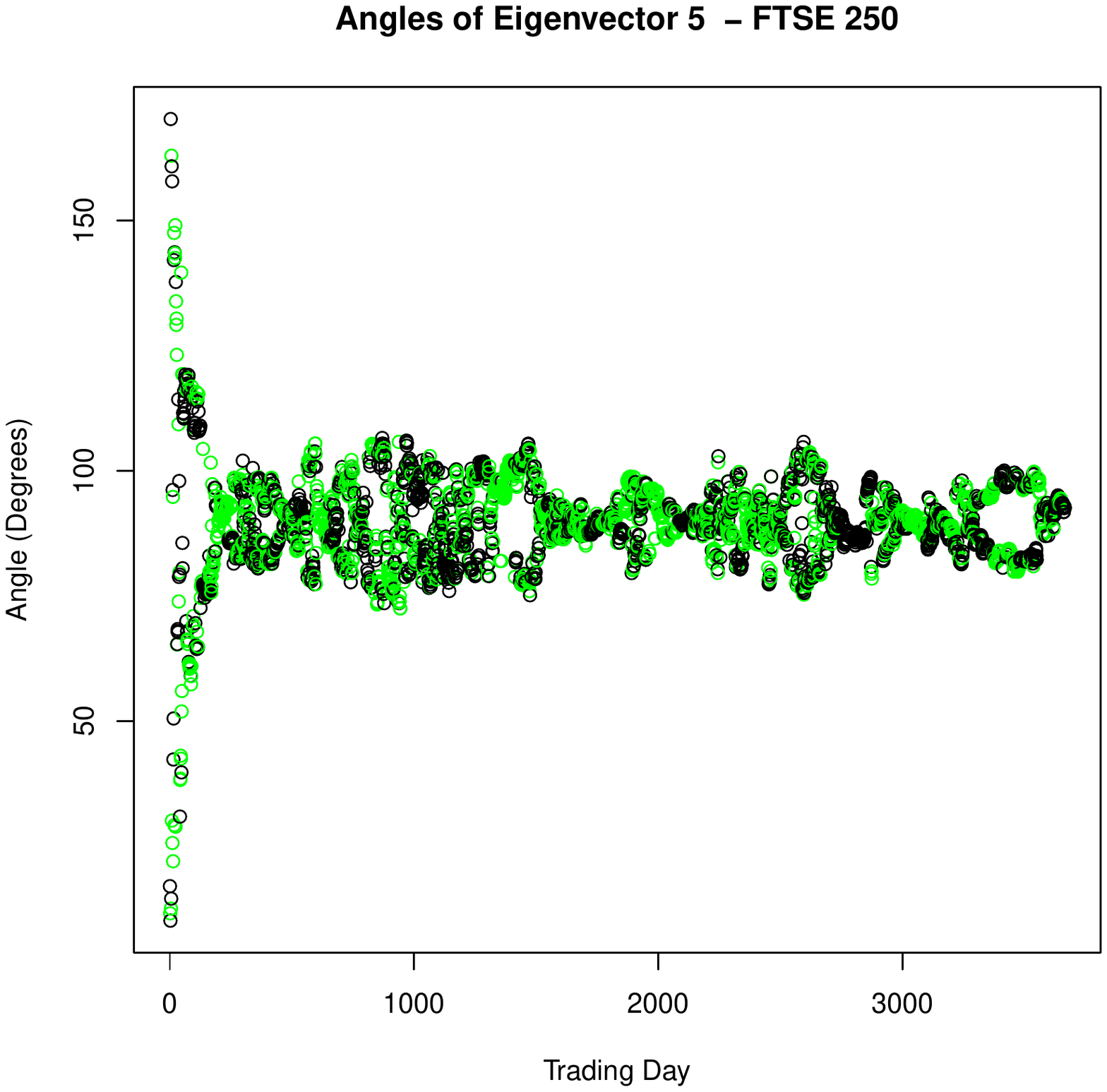}
  \caption{Angles of eigenvectors of PC 5.
 }
\label{fig:AnglesFTSE_PC5}
\end{figure}

\begin{figure}[ht]
  \centering
  \includegraphics[width=14cm]{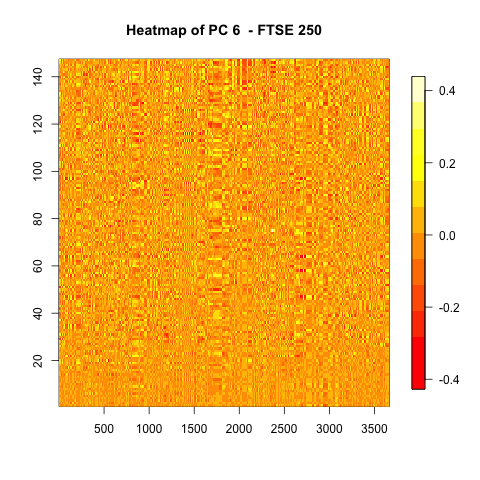}
  \caption{Heatmap of coeffients of PC 6.
 }
\label{fig:HeatFTSE_PC6}
\end{figure}

\begin{figure}[ht]
  \centering
  \includegraphics[width=14cm]{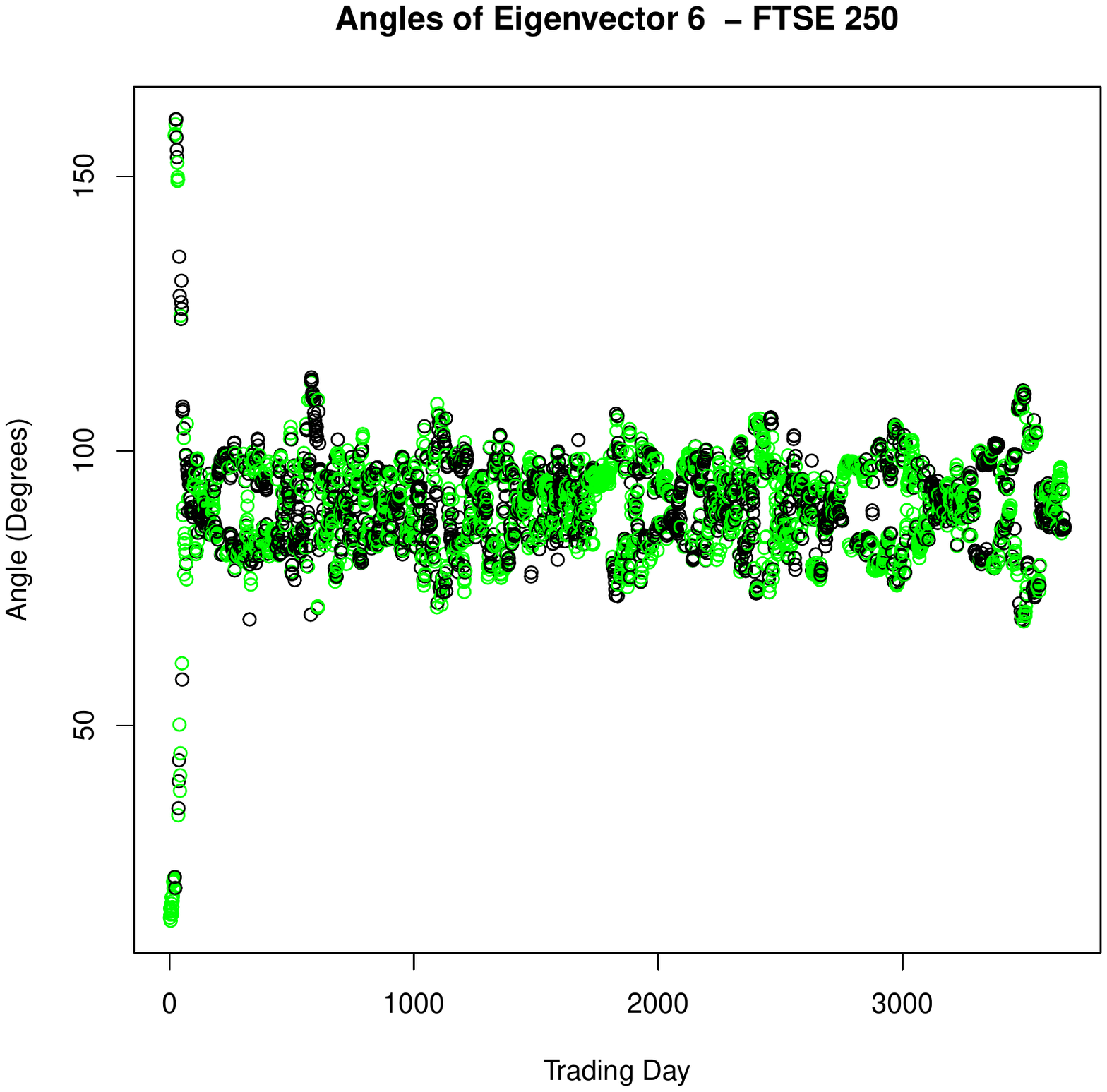}
  \caption{Angles of eigenvectors of PC 6.
 }
\label{fig:AnglesFTSE_PC6}
\end{figure}

\begin{figure}[ht]
  \centering
  \includegraphics[width=14cm]{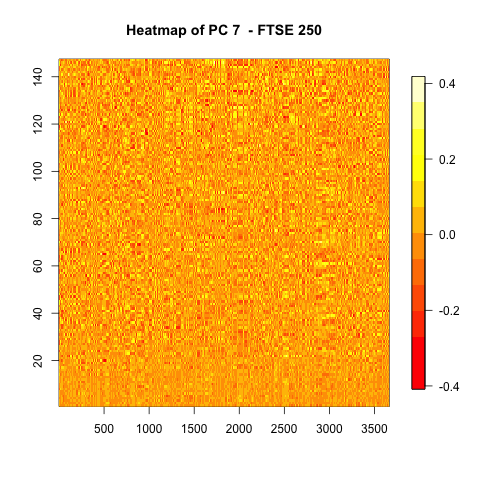}
  \caption{Heatmap of coeffients of PC 7.
 }
\label{fig:HeatFTSE_PC7}
\end{figure}

\begin{figure}[ht]
  \centering
  \includegraphics[width=14cm]{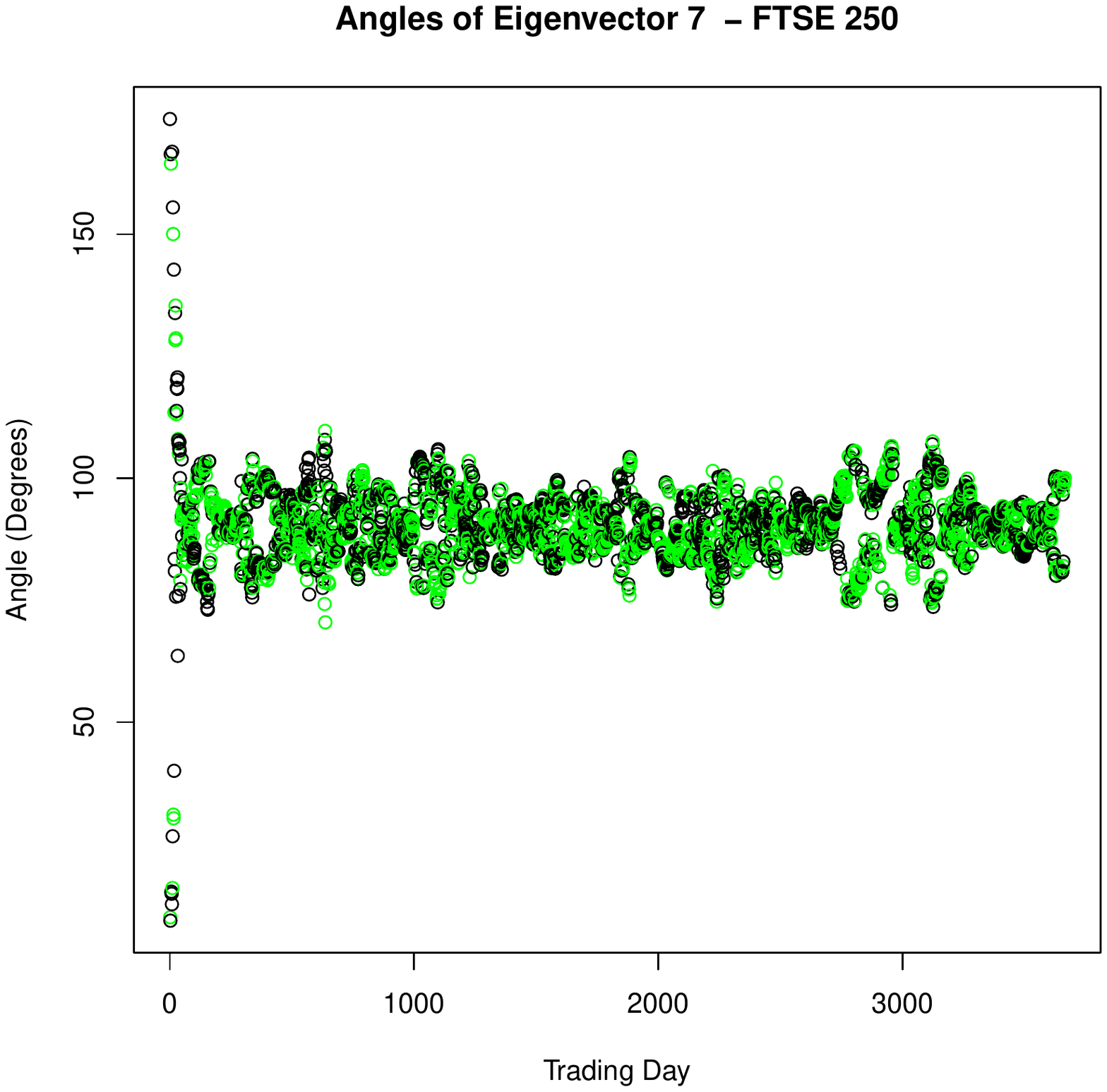}
  \caption{Angles of eigenvectors of PC 7.
 }
\label{fig:AnglesFTSE_PC7}
\end{figure}

\begin{figure}[ht]
  \centering
  \includegraphics[width=14cm]{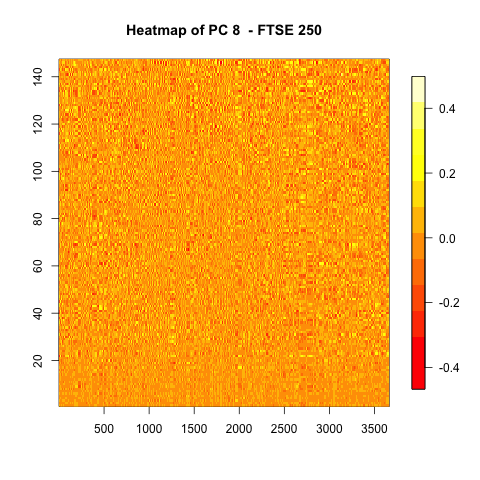}
  \caption{Heatmap of coeffients of PC 8.
 }
\label{fig:HeatFTSE_PC8}
\end{figure}

\begin{figure}[ht]
  \centering
  \includegraphics[width=14cm]{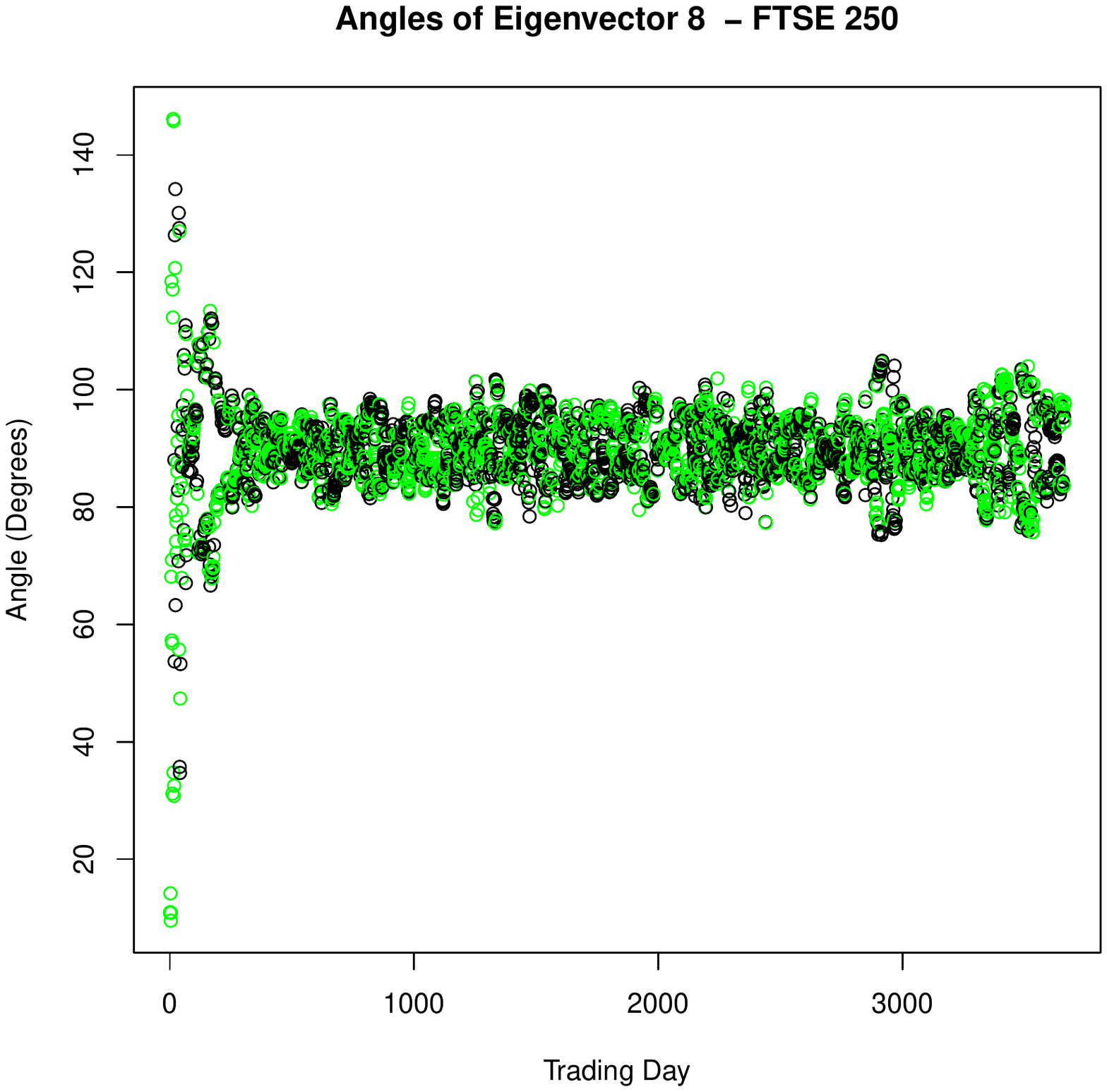}
  \caption{Angles of eigenvectors of PC 8.
 }
\label{fig:AnglesFTSE_PC8}
\end{figure}

\begin{figure}[ht]
  \centering
  \includegraphics[width=14cm]{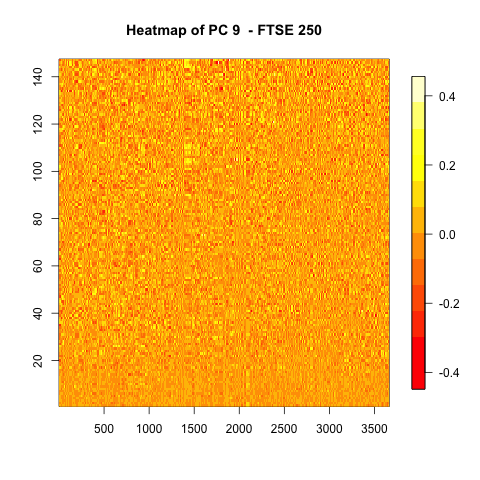}
  \caption{Heatmap of coeffients of PC 9.
 }
\label{fig:HeatFTSE_PC9}
\end{figure}

\begin{figure}[ht]
  \centering
  \includegraphics[width=14cm]{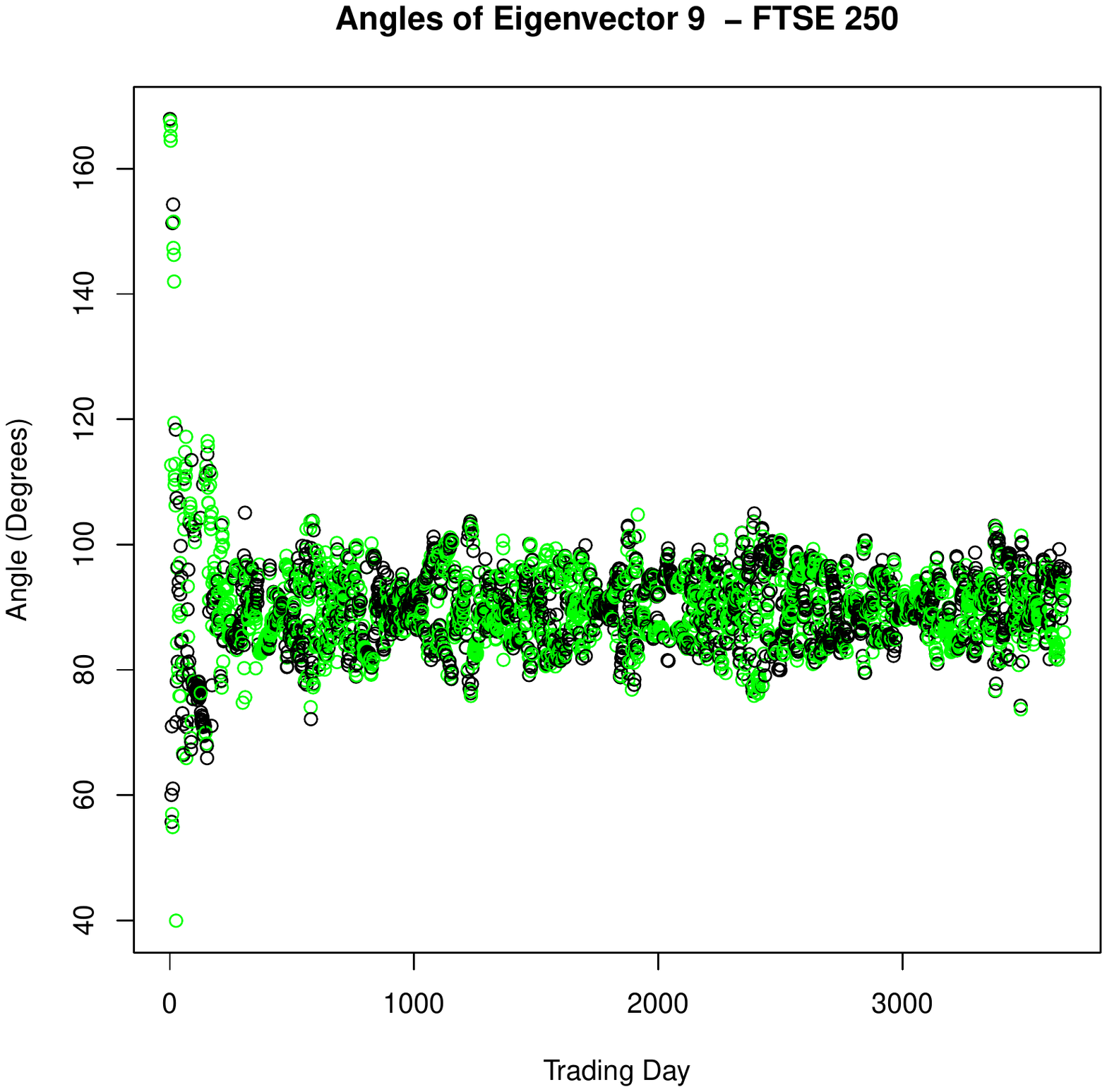}
  \caption{Angles of eigenvectors of PC 9.
 }
\label{fig:AnglesFTSE_PC9}
\end{figure}

\begin{figure}[ht]
  \centering
  \includegraphics[width=14cm]{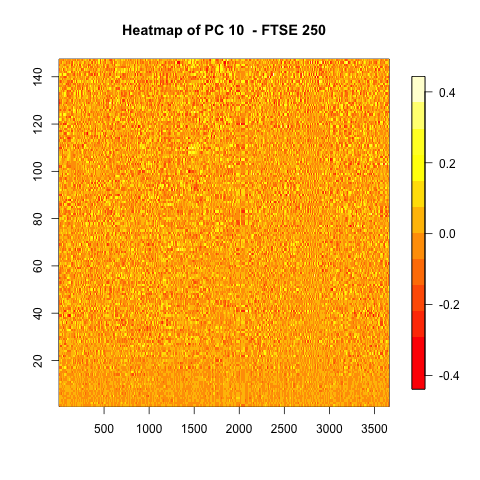}
  \caption{Heatmap of coeffients of PC 10.
 }
\label{fig:HeatFTSE_PC10}
\end{figure}

\begin{figure}[ht]
  \centering
 \includegraphics[width=14cm]{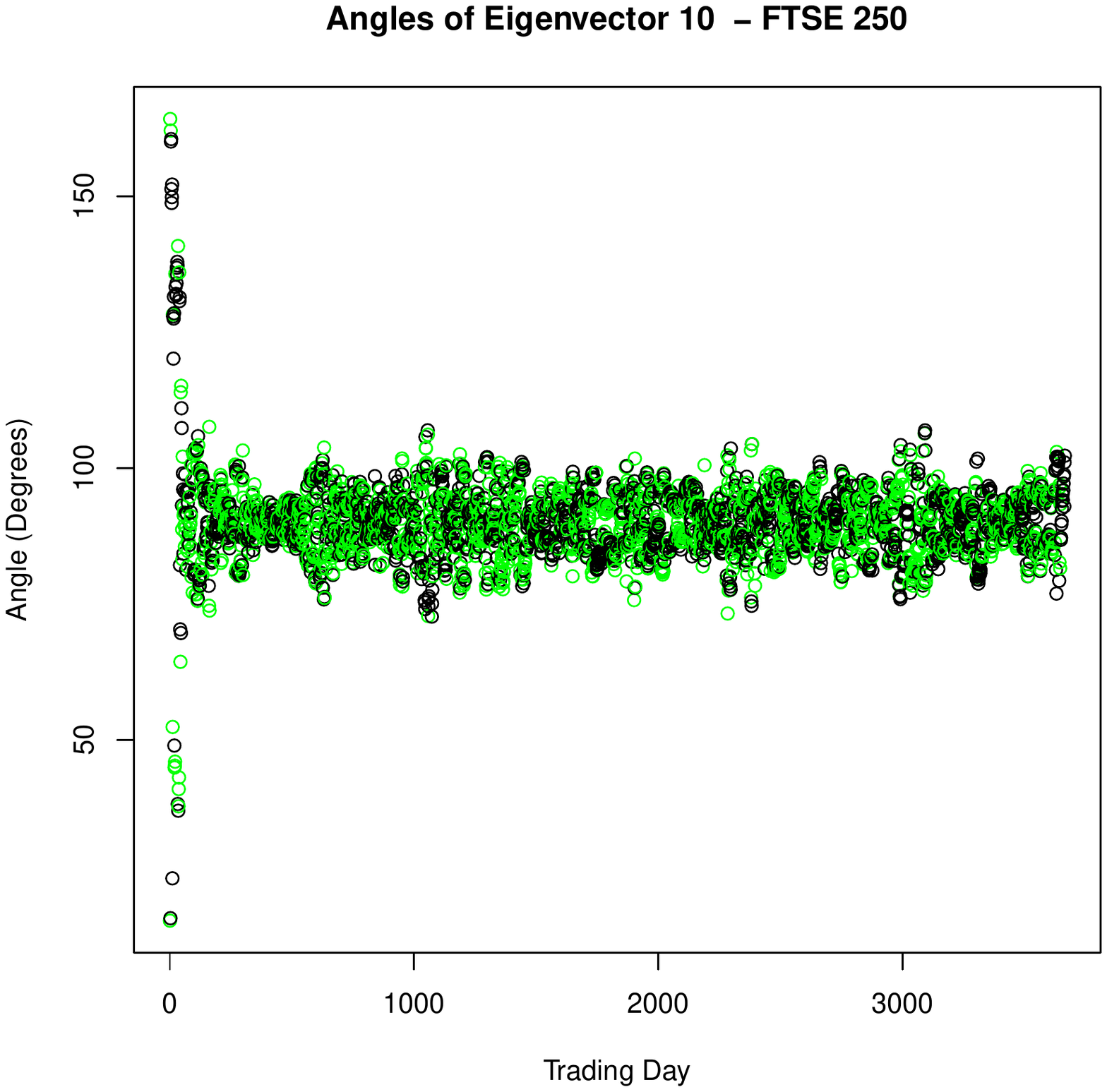}
  \caption{Angles of eigenvectors of PC 10.
 }
\label{fig:AnglesFTSE_PC10}
\end{figure}
\clearpage

\section{Additional Heatmaps and Angles -- Meteorology Data}

\begin{figure}[ht]
  \centering
  \includegraphics[width=14cm]{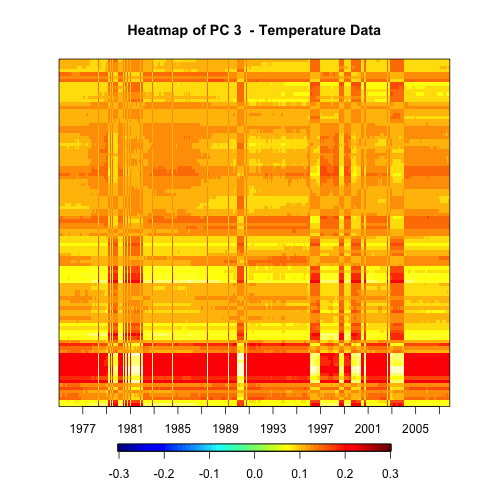}
  \caption{Heatmap of coeffients of PC 3.
 }
\label{fig:HeatBOM_PC3}
\end{figure}

\begin{figure}[ht]
  \centering
  \includegraphics[width=14cm]{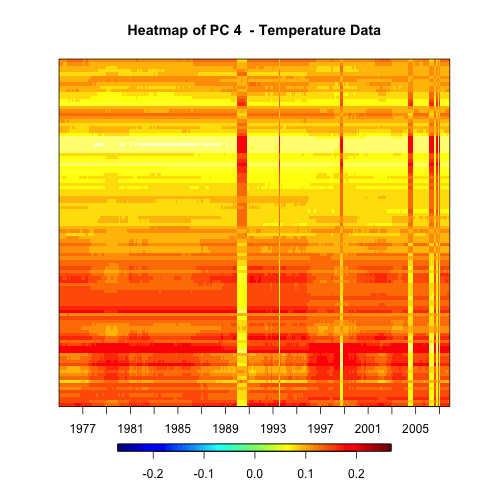}
  \caption{Heatmap of coeffients of PC 4.
 }
\label{fig:HeatBOM_PC4}
\end{figure}

\begin{figure}[ht]
  \centering
  \includegraphics[width=14cm]{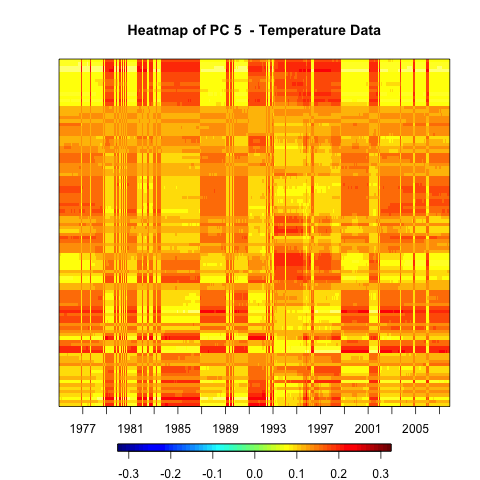}
  \caption{Heatmap of coeffients of PC 5.
 }
\label{fig:HeatBOM_PC5}
\end{figure}

\begin{figure}[ht]
  \centering
  \includegraphics[width=14cm]{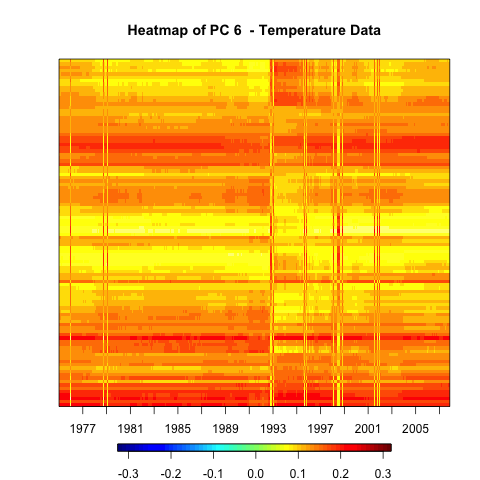}
  \caption{Heatmap of coeffients of PC 6.
 }
\label{fig:HeatBOM_PC6}
\end{figure}

\begin{figure}[ht]
  \centering
  \includegraphics[width=14cm]{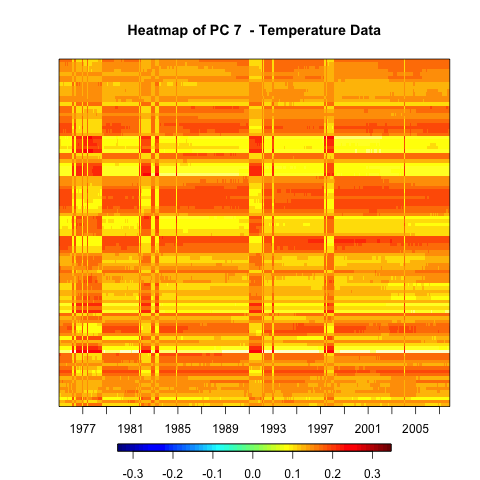}
  \caption{Heatmap of coeffients of PC 7.
 }
\label{fig:HeatBOM_PC7}
\end{figure}

%

\begin{figure}[ht]
  \centering
  \includegraphics[width=14cm]{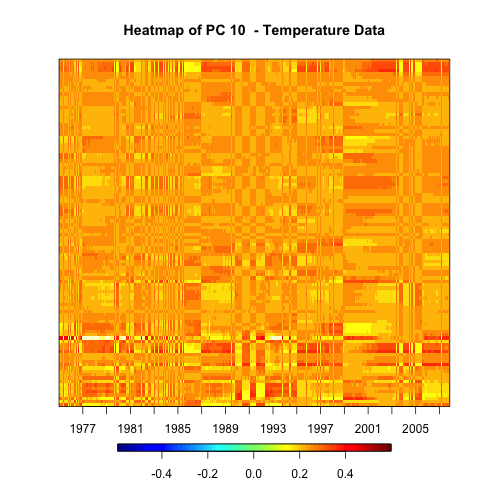}
  \caption{Heatmap of coeffients of PC 10.
 }
\label{fig:HeatBOM_PC10}
\end{figure}

\clearpage

\end{document}